\documentclass[trackchanges,twocolumn]{aastex6}

\usepackage{amsmath}
\usepackage{amsfonts}
\usepackage{amssymb}

\begin{document}


\title{Resolved star formation efficiency in the Antennae galaxies}


\author{Allison~M.~Matthews\altaffilmark{1}}
\author{Kelsey~E.~Johnson\altaffilmark{1}}
\author{Bradley~C.~Whitmore\altaffilmark{2}}
\author{Crystal~L.~Brogan\altaffilmark{3}}
\author{Adam K. Leroy\altaffilmark{4}}
\author{Remy Indebetouw\altaffilmark{1,3}}


\altaffiltext{1}{Department of Astronomy, University of Virginia, P.~O.~Box 400325 Charlottesville, VA 22904-4325, USA}
\altaffiltext{2}{Space Telescope Science Institute (STScI), 3700 San Martin Drive, Baltimore, MD 21218, USA}
\altaffiltext{3}{National Radio Astronomy Observatory, 520 Edgemont Rd., Charlottesville, VA 22903, USA 0000-0002-6558-7653}
\altaffiltext{4}{Department of Astronomy, The Ohio State University, 140 West 18th Avenue, Columbus, OH 43210, USA 0000-0002-2545-1700}

\begin{abstract}

We use Atacama Large Millimeter Array CO(3-2) observations in conjunction with optical observations from the {\it Hubble Space Telescope} to determine the ratio of stellar to gas mass for regions in the Antennae Galaxies.  We adopt the term ``instantaneous mass ratio''  IMR$(t)=M_{stars}/(M_{gas}+M_{stars})$, that is equivalent to the star formation efficiency for an idealized system at $t=0$.  We use two complementary approaches to determining the IMR$(t)$ based on 1) the enclosed stellar and molecular mass within circular apertures centered on optically-identified clusters, and 2) a tessellation algorithm that defines regions based on CO emission.  We find that only a small number of clusters appear to have IMR$(0)$~$=$~SFE~$>0.2$, which suggests that only a small fraction of these clusters will remain bound. The results suggest that by ages of 10$^{6.7}$~years, some clusters will have lost all of their associated molecular gas, and by 10$^{7.5}$~years this is true for the majority of clusters.  There appears to be slight dependence of the IMR$(t)$ on the CO surface brightness, which could support the idea that dense molecular environments are more likely to form bound clusters. However, the IMR$(t)$ appears to have a strong dependence on extinction, which likely traces the evolutionary state of clusters.  
 
\end{abstract}


\section{Introduction}
\label{sec:intro}
The efficiency with which stars form in a given environment is a fundamental parameter in our attempts to understand galaxy evolution throughout cosmic time.   For example, cosmological simulation codes must adopt values for the stellar mass that will form from a given region of gas over a specified time period \cite[e.g.][]{springel05, vogelsberger14}.  A significant body of work has been devoted to understanding how star formation efficiency (SFE) depends on various environmental parameters, relating gas phase and surface density to observed star formation rates (SFRs) \citep[e.g.][]{kennicutt89, bigiel08, leroy08}.  The set of factors that ultimately determine the SFE of a system remains an open issue, although work with increasingly sophisticated models is providing insight.  For example, using the Hyperion code, \citet{raskutti16} find that the SFE of a molecular cloud is strongly increased by turbulent compression of gas, which is, at least in principle, observationally testable.  

``Star formation efficiency'' (SFE) is defined in different ways depending on the astrophysical context.  In extragalactic research, the ratio of the star formation rate (SFR) to the total gas content (SFE=SFR/$M_{\rm H2}$) is frequently adopted \citep[e.g.][]{kirkpatrick17, li17, young1996}. This definition can also be thought of as an inverse depletion time.  Extragalactic studies frequently observe properties related to star formation and gas mass over relatively large spatial scales ($\sim 1$~kpc), which averages over star-forming regions during multiple phases of their evolution.  Another common modern definition of the star formation efficiency is the ratio of the gravitational free fall time to the depletion time. This expresses the fraction of gas mass converted into stars per gravitational free fall time. It is predicted by many theoretical models and the low values $\lesssim 1\%$ found indicate the inefficiency of star formation compared to free collapse (e.g., see \cite{krumholz2005}, \cite{mckee2007}, Federrath \& Klessen 2013). Studies that probe smaller size scales find that the general star formation ``laws'' derived from averaging over large scales (and the range of star formation stages) start to break down \citep[e.g.][]{schruba10}. 

On the other hand, studies of the Milky Way often resolve individual molecular clouds and typically define star formation efficiency as the fraction of gas mass that has been converted to stellar mass (SFE=$M_{stars}/(M_{gas}+M_{stars})$). In contrast to extragalactic work, these studies are not averaging over the full range of stages of star formation, but rather are necessarily focused on the early phases in which molecular material is still present. Thus, the aim with these resolved studies is not to derive average depletion times, but rather to determine the total mass in stars that results from a given amount of gas in a localized region.  These Galactic studies typically find values $\lesssim 30$\% for embedded Galactic clusters \citep[e.g.][]{lada03}. Subsequent work by \citet[][]{evans09} and \citet[][]{peterson11} derive even lower values of SFE, deriving values of $<10$\% for individual clouds in the Milky Way.  

Applying the methods typically used in Galactic studies to systems outside the Local Group has been challenging.  Due to observational limitations, studies of extragalactic SFE have typically been limited to nearby galaxies ($\lesssim 10$~Mpc), large spatial scales ($\gtrsim 500$~pc), or both.  Now with the availability of ALMA observations, it is possible to probe SFE on the same size scale as that of giant molecular clouds (GMCs) of $\lesssim 100$~pc \citep{schinnerer2013,leroy2016}.  Resolving size scales of $\lesssim 50$~pc is particularly important for probing the SFE involved in the formation of the so-called ``super star clusters'' (SSCs).  

\subsection{Super Star Cluster Evolution}
\label{sec:1.1}

SSCs are the most massive and dense type of stellar cluster, and represent one of the most extreme modes of star formation in the universe \citep[][]{oconnell94}.  First identified as bright blue knots of star formation \citep[e.g.][]{arp85, melnick85}, these objects are now generally considered to be the precursors to globular clusters \citep[][]{meurer95}.  However, various lines of evidence suggest that perhaps as few as $\sim 1$\% of SSCs survive to become globular clusters \citep[][]{fall06, fall05, fall01}. A number of factors play a role in determining whether a specific SSC will survive to become a globular cluster, including stellar feedback, tidal shocks, two-body relaxation, and tidal truncation \citep{gnedin97}.  One of the first criteria that a given SSC must meet to (eventually) evolve into a globular cluster is that it must be gravitationally bound.  Moreover, even if an SSC is originally bound at the time of its formation, if a significant amount of the binding mass is in the form of gas, the cluster is likely to become unbound when the gas is removed through stellar feedback. Thus, SFE also has an important role in the formation and survival of globular clusters. 

The impact of SFE on the evolution and long-term survival of star clusters has been studied by a number of authors for a distribution of conditions.  
For example, early work by \citet{hills80} based on the virial theorem suggested that some clusters can lose as little as 10\% of their mass (SFE $=90$\%) and still become unbound  after the gas was lost from the system. The general requirement for high SFEs is supported by \cite{geyer01}, who suggest that the formation of bound clusters (e.g., globular clusters) requires extremely high star formation efficiencies of $\sim 50$\%. 


More recent work using simulations has been able to study the impact of SFE on cluster evolution using more sophisticated physical processes.  
For example, using a grid of simulations \citet{baumgardt07} found that star clusters can remain bound with SFEs as low as 10\% provided that gas is removed from the clusters over long timescales. If, however, gas is removed instantaneously, clusters with SFE $<$ 33\% did not survive. Subsequent work by \citet{Pelupessy2012} used simulations of clusters with 1000 stars to explore their dynamical evolution under different scenarios.  Their simulations allow for clusters with SFEs as low as $5$\% to survive, but as with \citet{baumgardt07}, cluster survival is more likely if the gas removal does not take place during violent quasi-instantaneous episodes.

Observationally determining the SFE for globular clusters presents numerous challenges.  The first issue is that with current observing capabilities, it is not possible to observe the formation of globular clusters $\gtrsim 10$~Gyr in the past. Instead, we turn to relatively nearby objects that are generally considered to be analogous to adolescent globular clusters -- super star clusters.  The second issue is that determining the SFE requires knowing {\it both} the total stellar mass that forms in a given molecular cloud, and the total mass of the molecular cloud.  However, these two properties are largely mutually exclusive -- once stars have formed they will start disrupting the cloud, and if there is still molecular material available, additional stars could form. 
In fact, whether molecular material can remain or be reintroduced after the first generation of stars is being investigated by a number of studies.  For example, whether various types of ``polluters'' \citep[e.g. AGB stars, fast rotating massive stars (FRMS), and massive stars in binary systems][]{dercole08,renzini08,conroy11,decressin09,krause13,demink09} was explored by \cite{cabrera-ziri15} in three young (50-200\,Myr) molecular clouds. They found no evidence of CO(3-2) emission in these YMCs, placing strong constraints on the various ``polluter'' scenarios.

This problem can be at least partially mitigated if a statistically significant sample of SSCs with known ages is available.  In this case, the relative fraction of stellar and gas mass can be tracked through a cluster's evolution.  In this case, the ratio $M_{stars}/(M_{gas}+M_{stars})$ becomes a function of time, and here we adopt the term ``instantaneous mass ratio'' (IMR).  The IMR is only equal to the actual SFE at a time when stars have ceased forming and the molecular cloud has not been disrupted.  One can debate whether such a time is overly idealized or actually exists  -- whether stars have truly stopped forming, or whether the molecular cloud is not significantly disrupted until star formation has ceased.  In any case, one would predict a trend in which the youngest SSCs would be associated with the lowest IMR values, and that as the molecular cloud is destroyed the IMR should approach unity.   

\subsection{The Antennae}

In order to constrain the impact of SFE on cluster survival through observations, an optimal target will be relatively nearby (for both sensitivity and resolution limitations), and also host a statistical sample of star clusters.   
As the closest and youngest example in Toomre's 1977 sequence of prototypical mergers \citep{toomre1977}, the Antennae galaxies (NGC 4038/39) have been a target for multiwavelength observations on the internal structure of galaxy mergers. The IRAS satellite found that virtually all extragalactic luminous infrared sources were mergers \citep{sanders1988}, making them a prime laboratory for studying early stages of star formation. With an infrared luminosity of $L_{IR}\sim 10^{11} \, L_{\odot}$, the Antennae galaxy system is barely classified as a LIRG, but the vast molecular gas reservoir \citep[$>10^{10} \, M_{\odot}$,][]{gao2001, Wilson2003} suggests that the Antennae may evolve into a superluminous infrared galaxy as the merger continues.

With a molecular gas reservoir comparable to that of an early-stage ULIRG, the Antennae has an abundance of the ingredients needed for star formation. HST observations have identified $>10,000$ stellar clusters in the overlap region alone \citep{whitmore10}. The close distance to the Antennae galaxies \citep[$d\sim22$ \, Mpc,][]{schweizer2008} yields better linear resolution than available for ULIRGs more common at higher redshifts. As a nearby snapshot in the evolution of a gas-rich merger, the Antennae provides a crucial opportunity to better understand when, and how efficiently stars form in a merger environment and the initiation of starburst galaxies in exquisite detail.

The organizational structure of this paper is as follows. In Section \ref{sec:observations} we describe the sub-mm and optical observations utilized in this study and methods used for calculating the stellar and molecular masses. In Section \ref{sec:stellar} we present the process of determining the IMR$(t)$ through measurements centered on stellar clusters. In Section \ref{sec:co} we present the algorithm and method for determining the IMR$(t)$ through measurements centered on molecular gas. In Section \ref{sec:results} we report the total mass information and instantaneous mass ratios for both measurement schemes. In Section \ref{sec:discussion} we discuss the implications of the results in the context of understanding early stages of star formation. Finally, in Section \ref{sec:conclusions} we summarize our findings.

\section{Observations and Methods}
\label{sec:observations}
\subsection{ALMA Observations}
\label{sec:alma}

\begin{figure}
    \centering
    \includegraphics[trim={2.7cm 0cm 3.3cm 0cm},clip,width=0.48\textwidth]{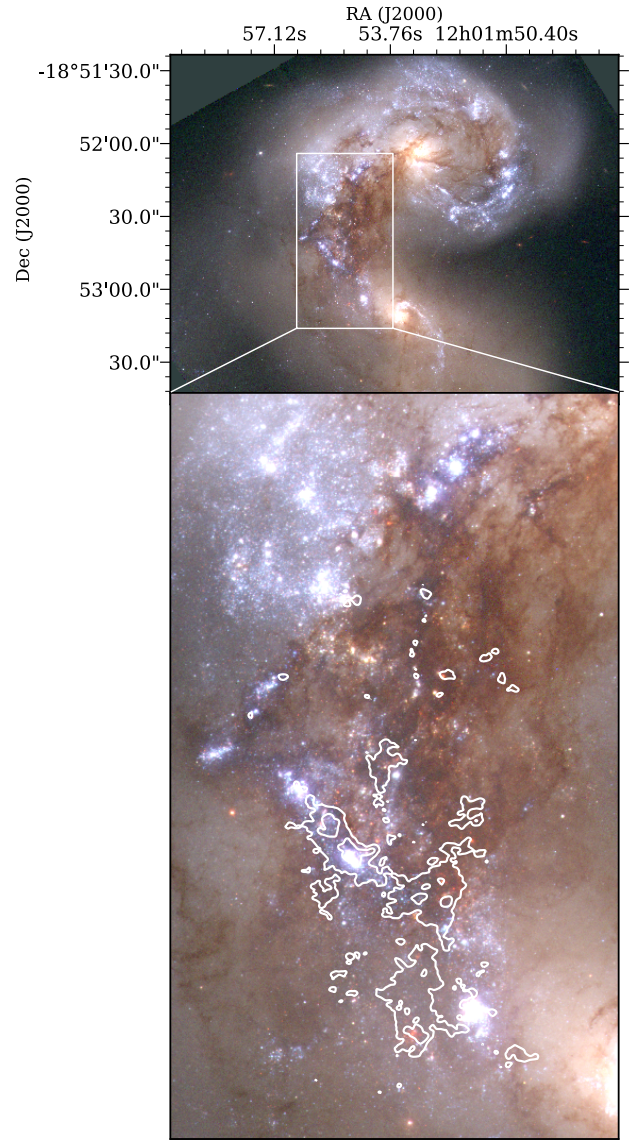}
    \caption{An optical image of the Antennae taken with {\it HST} is shown in the top panel. The white box shows the approximate region covered by the ALMA Cycle 0 observations. This region is expanded in the bottom panel with CO(3-2) contours overlaid in white. The contour levels are $[55, 400]$\,K\,km/s.}
    \label{fig:antennaeWhole}
\end{figure}

We utilize previously attained ALMA CO(3-2) Cycle~0 data of the Antennae overlap region (Figure \ref{fig:antennaeWhole}). These data were first presented in \cite{whitmore14}, and detailed information on the configuration and calibration of such data is given in that work. In summary, a 13 point mosaic was obtained with the ``extended'' configuration ($\sim$~400\,m maximum) covering the overlap region between the two interacting galaxies. These data did not contain short or zero spacings, however the size-scales of the structures of interest here are sufficiently compact ($\lesssim 100$~pc $\approx 1"$) this is not a concern. 

The resulting CO~(3~-~2) data cube has an angular resolution FWHM of $0\farcs56 \times 0\farcs43$, equivalent to ($59 \times 45$ pc) at the distance of the Antennae, $\sim 22$ Mpc. The rms of this data cube was determined to be 3.3 mJy/beam using line-free channels. These observations enabled the high resolution co-spatial study of the mass of molecular gas as compared to the mass of stars and star clusters (i.e., the star formation efficiency). 

\subsection{HST Observations}
\label{sec:hst}

We retrieved {\it Hubble Space Telescope} images of the the Antennae galaxies (NGC 4038/4039) using the Advanced Camera for Surveys (ACS) from the Hubble Legacy Archive. The original observations were published in \cite{whitmore10} and include multi-band photometry in the following optical broadband filters: {\it F435W}, {\it F550M}, and {\it F814W}. All optical observations were made with the Wide Field Camera (WFC) on the ACS and exposure times total 2192~s, 2544~s, and 1680~s, respectively. The observations were taken with a scale of $0\farcs049$ pixel$^{-1}$ and yield a field of view of $\sim202''\times202''$. The registration of the HST optical images was done following the methods of \cite{brogan2010}. Catalogs of clusters and their optical properties in the Antennae were constructed from the $F550M$ image using DAOFIND in IRAF and were presented in \cite{whitmore10}. For our purposes, we use the ``concentration index'' $CI$ of each source in the catalog to select for clusters and remove any bright stars from our sample. This is defined in \cite{whitmore10} as the different between magnitudes measured within a 1 and a 3 pixel radius with more concentrated objects have lower values of $CI$. As in \cite{whitmore10} we consider sources with $CI>1.52$ to be clusters and are included in our analysis.

\subsection{Determining Mass-to-light ratios}
\label{sec:m/l}

Our prescription for the IMR includes a term representing the stellar mass of a given region, which therefore necessitates determining the mass-to-light ratio, ${\rm M}/L_{814}$. We attempted several methods of determining the mass-to-light ratio of spatial bins produced by the Weighted Voronoi Tessellation (WVT, Section \ref{sec:co}) before settling on the procedure described later in this section. These approaches aimed to characterize the mass-to-light ratio in the Antennae overlap region as a function of age using 1) model predictions and 2) an empirical relationship.

For the first approach, we generated Starburst99 models for the I-band mass-to-light ratio, ${\rm M}/L_I$, i.e. we generated models for $M_V$ and $V-I$ and combined them in such a way to result in $M_I$ \citep{s99}. These models were generated for a total cluster masses of $10^4 - 10^7 {\rm M_{\odot}}$, various metallicity models, $Z$ from $0.008-0.04$, Salpeter and Kroupa IMFs, and upper mass limits on the IMF from $20-100 {\rm M_{\odot}}$. From these models, we generated functions of ${\rm M}/L_{I}$ with respect to age. The reduced chi-squared statistic of the models was calculated with respect to the relationship of cluster ${\rm M}/L_{I}$ with age derived from \cite{whitmore10}. In addition to chi-square values $>100$, adopting this approach requires assuming an extinction value {\it a priori}, and given the range of extinction values (particularly for clusters with ages $\lesssim 10$~Myr), we found that this method was not robust. 


We then attempted to derive a strictly empirical relationship between the observed value of cluster ${\rm M}/L_I$ from \cite{whitmore10} as a function of age. This was done by calculating the median of cluster ${\rm M}/L_I$ for each logarithmic age bin as described below. 

\begin{figure*}
\gridline{\fig{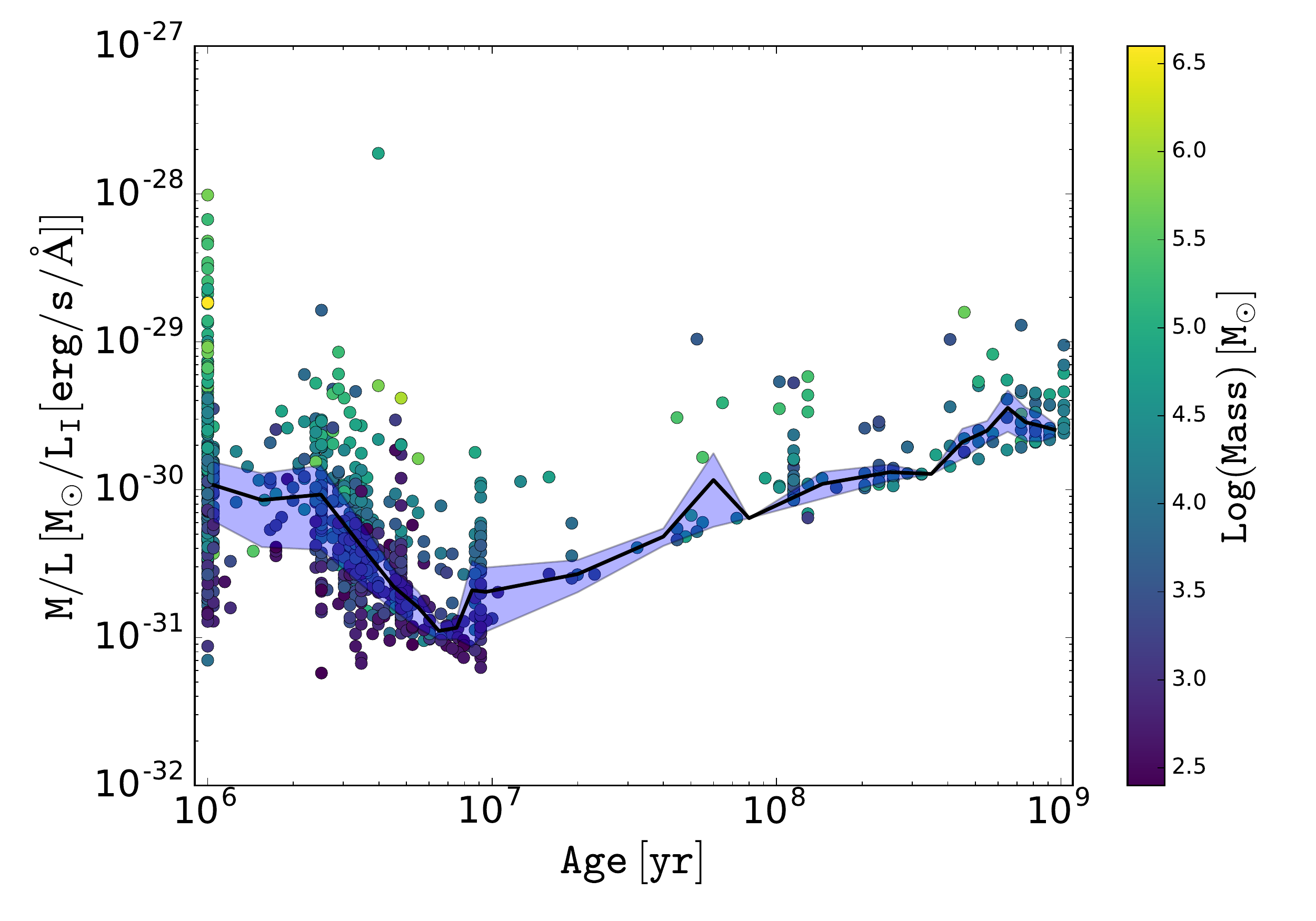}{0.49\textwidth}{(a)}
\fig{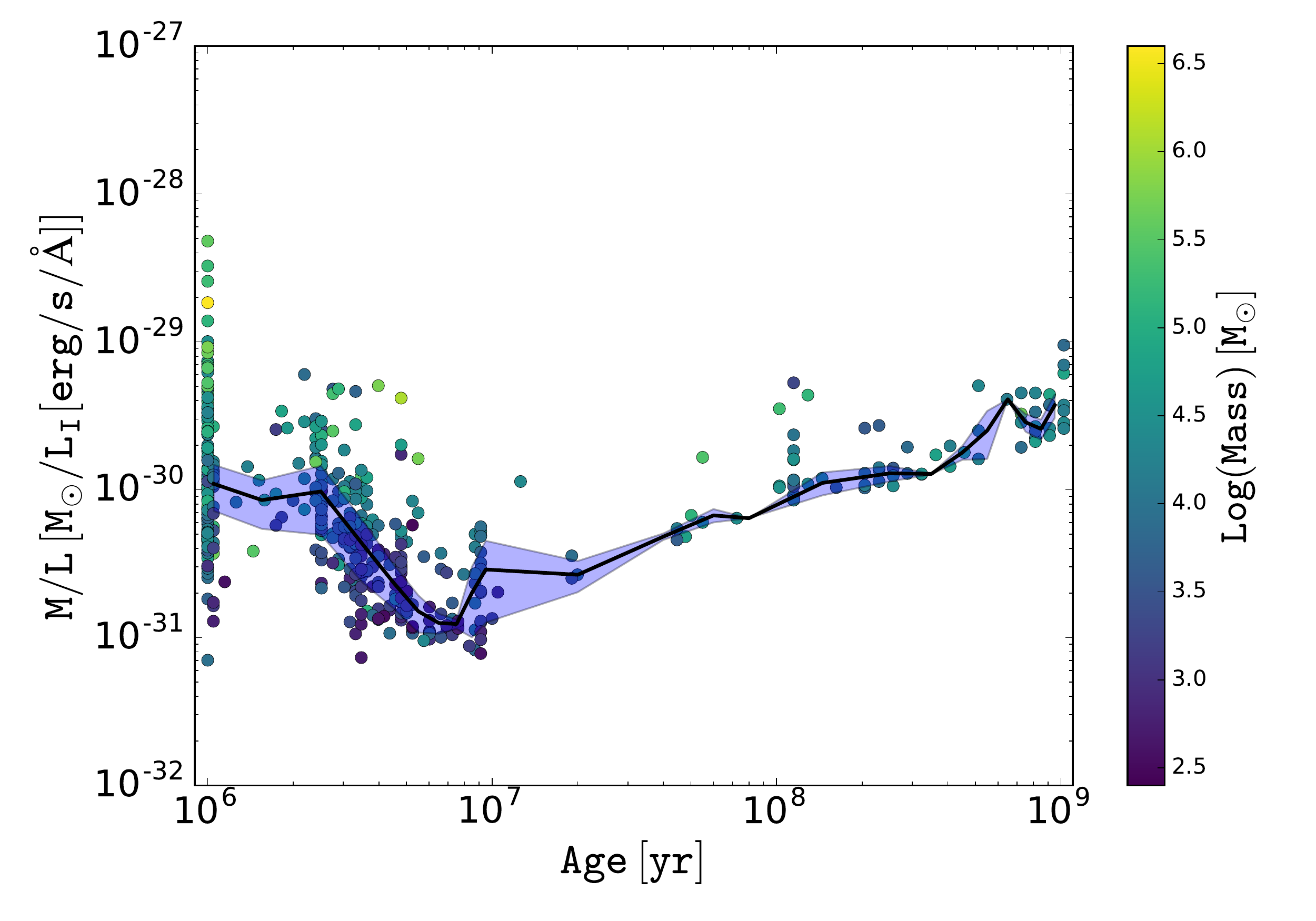}{0.49\textwidth}{(b)}}
\caption{The I-band mass-to-light ratio ($M/L_I$) in solar units taken from the measured cluster mass and I-band magnitudes of \cite{whitmore10} plotted versus age of the cluster. The solid black line indicates the median per logarithmic bin. The purple shaded region shows the absolute median deviation around the median value per logarithmic bin. Subfigure (a) includes all clusters in the Antennae overlap region observed by ALMA Cycle 0. Subfigure (b) includes all clusters with concentration index $CI > 1.52$. 
\label{fig:mlmedian}}
\end{figure*}

From \cite{whitmore10}, we gathered the following parameters for each cluster in the overlap region: mass ($\rm M_{\odot}$), age ($\tau$), apparent magnitude in the I-band ($\rm m_I$), and the best estimate of reddening ($\rm E(B-V)$). 
The I-band apparent magnitudes were then corrected for extinction using the procedure described in Section \ref{sec:wvtstellarmass} and converted to absolute magnitudes assuming a distance to the Antennae galaxies of $d=22$~Mpc. We then calculated the mass-to-light ratio to be the ratio of the mass to the corrected I-band luminosity for a given cluster. Figure \ref{fig:mlmedian} shows the relationship of this ratio versus age of the cluster.
Here we opt to use the I-band luminosity as the primary value for ``light'' as opposed to converting to a luminosity in solar luminosity units $L_{\odot}$, which requires additional assumptions in the conversion to bolometric luminosity.

As with the model-based approach, the scatter in the data at the youngest ages (the clusters we are most interested in) spans three decades in the mass-to-light ratio, and any ensemble approach improperly estimates ${\rm M}/L_I$ for the majority of the young clusters. For this reason, we viewed the ensemble median ${\rm M}/L_I$ as a function of age to be too general for the wide variety of clusters and environments. Rather, we assigned to each region the mass-to-light ratio of the enclosed star cluster. In the case where there are multiple clusters in a region, we assigned the median of those to that particular region. These mass-to-light ratios are shown in relation to positions in the Antennae overlap region in Figure \ref{fig:mltessel}.



\begin{figure}
    \centering
    \includegraphics[width=0.47\textwidth]{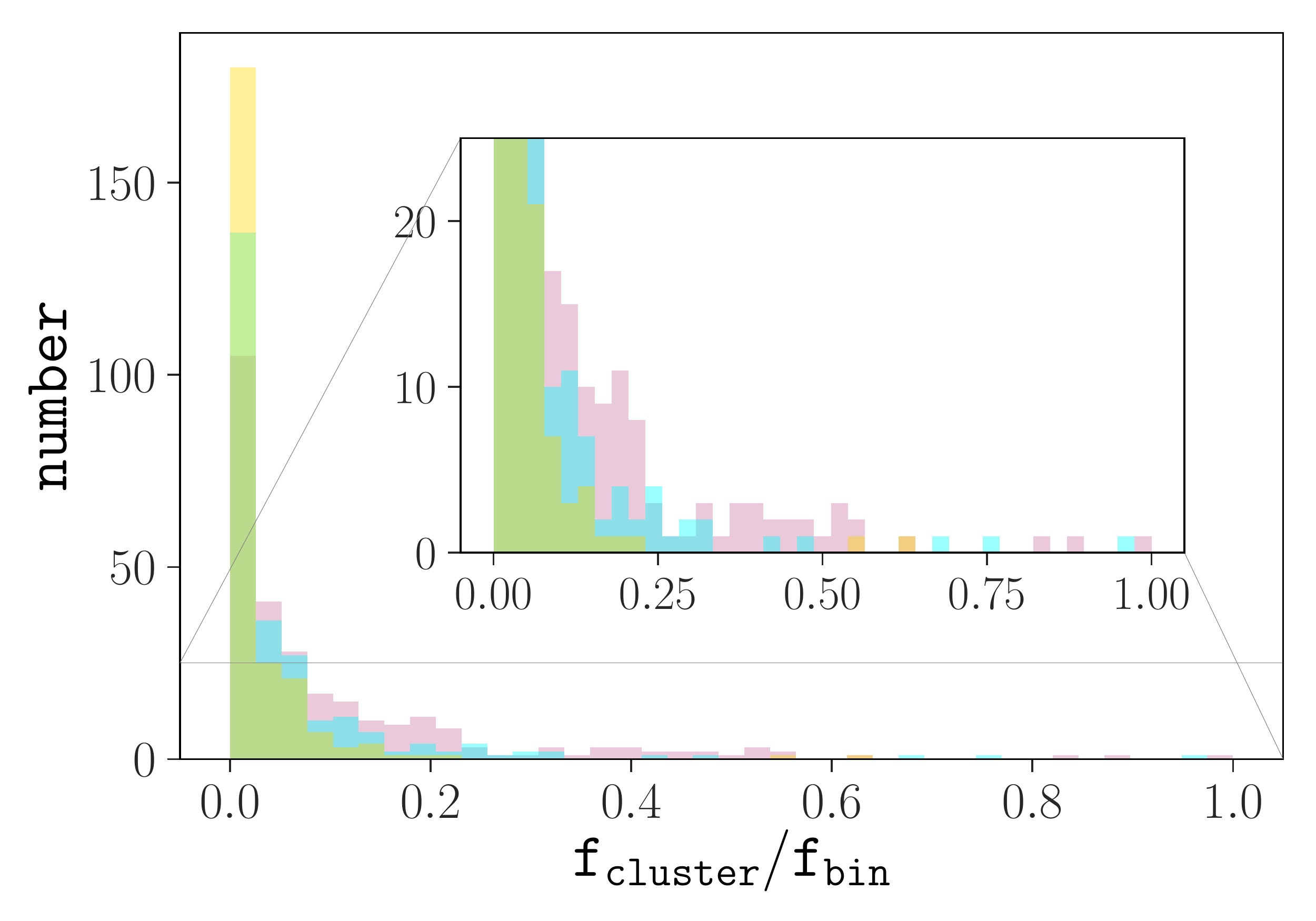}
    \caption{Distribution of fraction of light in cluster versus the light in the entire surrounding WVT bin. Inset shown to highlight the WVT bins with large fractions of light coming from the enclosed cluster. The various colors show the distributions for various WVT bin sizes: $L_{target} = 3.6 \times 10^5$\,K\,km/s\,pc$^2$ (mauve), $L_{target} = 7.3 \times 10^5$\,K\,km/s\,pc$^2$ (cyan), and $L_{target} = 14.5 \times 10^5$\,K\,km/s\,pc$^2$ (gold).}
    \label{fig:light_ratio}
\end{figure}

The assumptions made in this process regarding non-cluster light in a region follow those discussed in Section~\ref{sec:stellar}.  Namely, we assume that the diffuse inter-cluster light in the aperture between the cluster and the larger aperture boundary has approximately the same mass-to-light ratio as the central cluster, i.e. they are both dominated by the same population. In a given WVT bin that contains at least one star cluster, the mean fraction of light than can be attributed to the cluster is 11\%, the distribution of this fraction for all WVT bins with a star cluster can be seen in Figure \ref{fig:light_ratio}.

\subsection{Calculation of Gas Mass from CO Emission}
\label{sec:comass}
The total CO(3-2) emission in a given aperture was measured from the moment~0 (integrated intensity) map of the ALMA Cycle-0 observations (see Section \ref{sec:alma}). The $\alpha_{\rm CO}$ conversion factor assumes a rotational transition of CO(J=1$-$0). Since our measured emission of CO is the J=3$-$2 rotational transition, we must first scale our luminosity of this upper transition to that expected for the ground state. We adopt CO(J=1$-$0)~$\sim 2\times$~CO(J=3$-$2), consistent with the median value found by \cite{wilson08} for a sample of LIRGs and with the values derived by \cite{ueda2012} found for giant complexes in the Antennae. 




We assume a Milky Way metallicity for the Antennae galaxies. For this assumption, the CO-to-H$_2$ conversion factor is $4.3$ M$_{\odot}$ (K~km~s$^{-1}$~pc$^{2}$)$^{-1}$ \citep{Bolatto2013}.  The units of the $\alpha_{\rm CO}$ conversion factor necessitate a CO luminosity $L_{\rm CO}$, yet simply summing the flux density in a region yields a measurement in units K\,km\,s$^{-1}$\,pixel$^2$. We adopted the following to calculate the corresponding CO luminosity of a region. The pixel scale was taken from the header of the CO(J=3$-$2) moment-0 map as $\Delta=2.22\times10^{-5}$ deg, corresponding to 0\farcs08 pixels. Assuming a distance to the Antennae of $d = 22$ Mpc \citep{schweizer2008}, we determined the physical scale corresponding to a pixel, $~8.5$\,pc/pixel. We then converted the the summed CO(J=3$-$2) emission in a region with units K\,km\,s$^{-1}$\,pixel$^2$ to a CO luminosity by multiplying this total by the conversion factor 72.6\,pc$^2$/pixel$^2$. With the CO luminosity determined, we arrive at the following conversion from CO(J=3$-$2) to a molecular gas mass in solar units:
\begin{equation}
M_{mol} = 2~\alpha_{\rm CO}~L_{\rm CO}.
\end{equation}

\section{Determination of the IMR based on locations of stellar clusters}
\label{sec:stellar}

In order to determine the instantaneous mass ratio (IMR) for a statistical sample of SSCs, we utilize the ALMA and {\it HST} observations discussed in Sections~\ref{sec:alma} and \ref{sec:hst}. We base the identifications of regions on the locations of known SSCs, and determine the stellar and gas mass in the surrounding area. To determine the IMR of SSCs in the Antennae, we utilize the catalog of star clusters in \citet{whitmore10}. This method is fundamentally different from that used below in Section~\ref{sec:co} in that the IMR values are determined for local regions centered on specific star clusters, as opposed to being based on regions of molecular emission.  Each of these approaches will be subject to different biases. In order to use this cluster-based method, we needed to determine both the total stellar mass and total molecular mass within an aperture around each cluster. 

To determine the stellar mass in a given aperture, we used the following method.  First, clusters from the ``50 Most Luminous'' and ``50 Most Massive'' tables (Tables~6 and 7 of that paper) with ages $< 10^8$~years were identified that were contained within the footprint of the ALMA observations. The upper age limit of 10$^8$~years is motivated by the mass loss and mechanical luminosity evolution, which is expected to decrease abruptly at $\sim 30$~million years, and roughly level off by 100~million years \citep{leitherer99}. Although, the observed age distributions of clusters appear to be more complicated and several other effects may be coming into play \citep{fall12}.
The ages, masses, and E(B-V) values for these clusters were retrieved from the \citet{whitmore10} catalog.  
Using the data from the F814W filter, for each cluster we determined the stellar mass per electrons per second ($e^-$/s), which incorporated both the extinction and aperture corrections of \cite{whitmore10}.  These values of stellar mass per $e^-$/s derived from a specific cluster were then applied to the total F814W emission for a given aperture to determine the enclosed stellar mass.  This process was carried out for circular apertures of both 25~pc and 50~pc, which is well-matched to the synthesized beam of the ALMA observations (Figure \ref{fig:clusterZoom}).  The Hubble Observations with a pixel size of $0\farcs05$ ($= 5.33$~pc) are well-resolved at this scale. 

There were key assumptions that accompany this determination of the mass-to-light ratio within an aperture. Namely, we assume that the diffuse inter-cluster light has approximately the same mass-to-light ratio as the nearest cluster. 
If the non-cluster light in a given aperture has a significantly different age or extinction than that of the cluster, the adopted mass-to-light ratio will be skewed.  Based on Starburst99 models of cluster magnitudes as a function of age \citep{s99}, we can constrain the extent to which the adopted mass-to-light ratio is uncertain.  For example, if the ambient light in a given aperture is due to a stellar population that is an order of magnitude older than the constituent cluster, the adopted mass-to-light ratio based on the cluster will be roughly 2.5$\times$ lower than the value appropriate for the surrounding stellar population. 
The offset in the mass-to-light ratio due to age is likely to be partially offset by differences in extinction for a surrounding stellar population with a different age than the cluster. The younger a stellar population is, the more significant the extinction of its light is likely to be, and therefore the higher the adopted mass-to-light ratio for the cluster.  For the clusters used in this study, the mean inferred extinction is $A_{814}=0.55$ (the maximum inferred extinction is $A_{814}=5.3$).  If the light from the surrounding stellar population is subject to negligible extinction, the adopted mass-to-light ratio would be roughly 1.6$\times$ higher than the value appropriate for the surrounding stellar population, significantly offsetting the effect due to age.
In order to be conservative, here (and below in Section~\ref{sec:co}) we adopt the factor of 2.5 as the uncertainty in inferring the mass-to-light ratio of an aperture based on the age of the star cluster(s) present.  

The stellar mass in an aperture was combined with the molecular mass as follows to yield the IMR for these regions centered on stellar clusters:
\begin{equation}\label{eq:imr}
    {\rm IMR} = \frac{M_{stars}}{(M_{stars}+M_{gas})}.
\end{equation}

\begin{figure}
  \centering
  \includegraphics[trim={3.23cm 0cm 1.05cm 0cm},clip,width=0.47\textwidth]{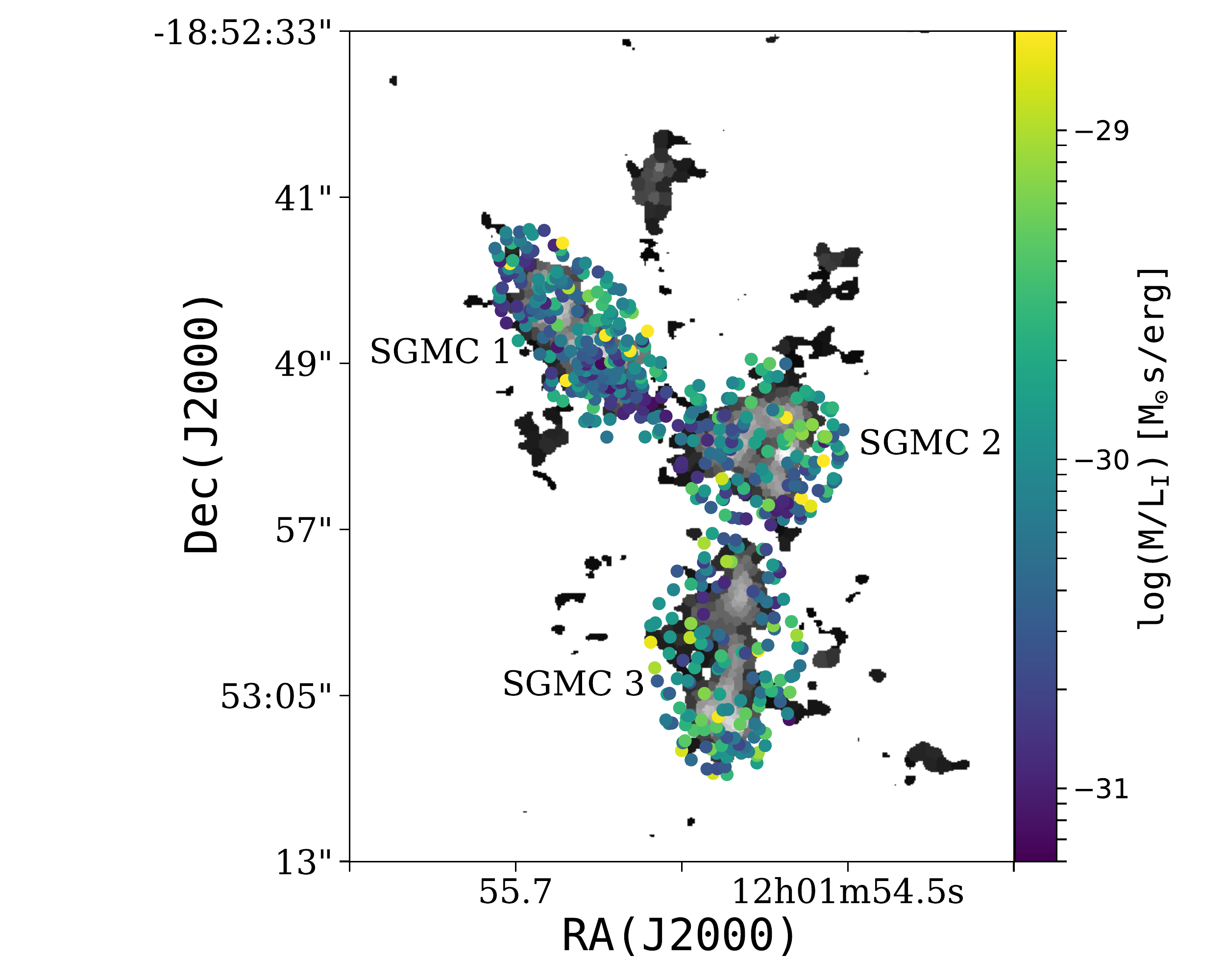}
\caption{The tessellation map for a target flux of $L_{target} = 7.3 \times 10^5$\,K\,km/s\,pc$^2$ of the Antennae overlap region is shown in grayscale. Overlaid are the mass-to-light ratios of the star clusters identified by \cite{whitmore10} in units of cluster mass in $M_{\odot}$ over cluster I band luminosity in $L_{\odot,I}$ for all clusters with concentration index, $CI > 1.52$.} \label{fig:mltessel}
\end{figure}

\section{Determination of the IMR based on locations of CO emission}
\label{sec:co}
\begin{figure*}
  \centering
  \includegraphics[scale=0.45,trim={1.9cm 0cm 2cm 0cm}, clip]{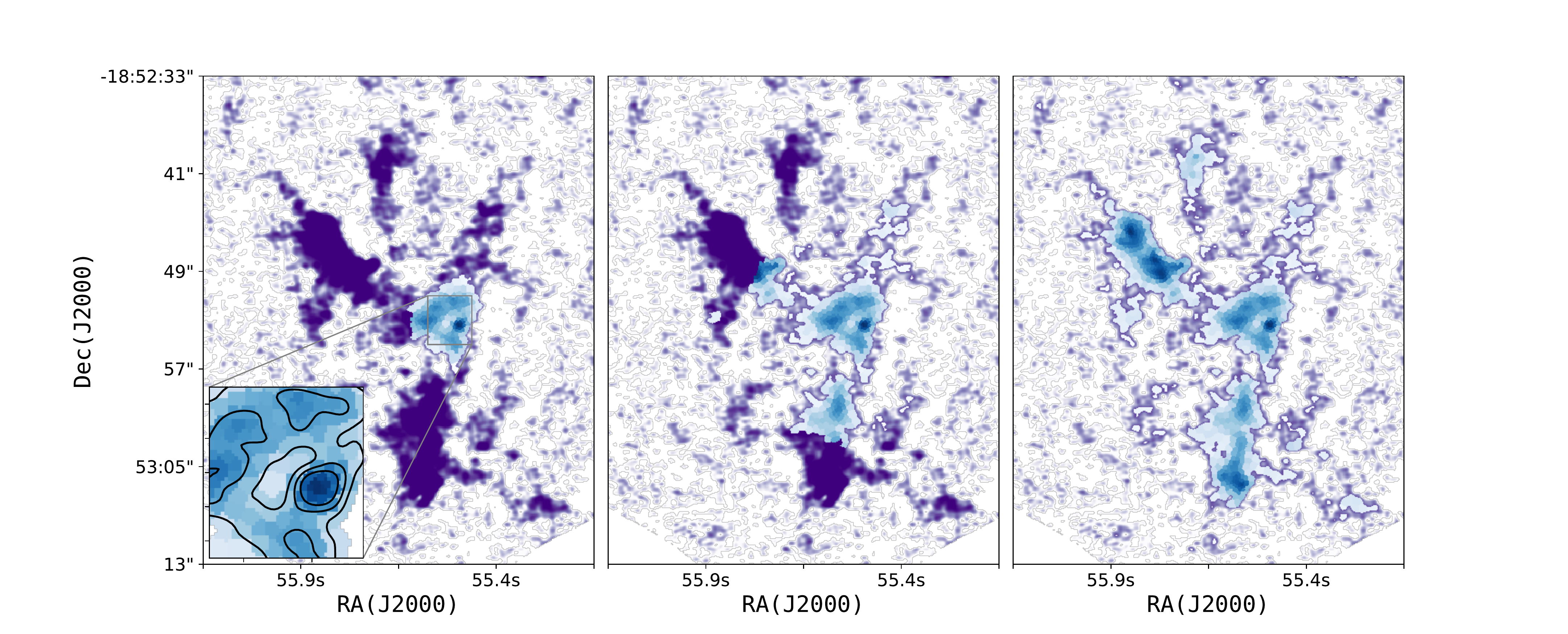}
  \caption{Timestep through the WVT binning algorithm of the Antennae overlap region shown in the bottom panel of Figure \ref{fig:antennaeWhole} after it has finalized 25\%, 50\%, and 100\% of the bins for a target luminosity of $3.6 \times 10^5$\,K\,km/s\,pc$^2$. Black contours represent the CO(3-2) emission with levels [165, 330, 500]\,K\,km\,s$^{-1}$.\label{fig:tesselTimeline}}
\end{figure*}

We then tried a more objective approach to determining the IMR in the overlap region of the Antennae galaxy system. Interest in the SFE at the most nascent stage of cluster formation lends itself well to identifying regions based upon the CO(3-2) integrated intensity map. Due to the complexity of the diffuse gas in the overlap region, we chose to adopt an algorithmic approach in defining these regions. While more traditional methods have involved identifying pockets of CO emission ``by eye,'' we believe this introduces biases due to the subjective opinions of differing authors. Our goal was to derive an objective collection of ``pixels'' (hereafter referred to as bins) such that the size and shape of the bins traces the density dependence of the emission from the gas. Our criteria for the algorithm did not include the ability to distinguish individual molecular clouds, and therefore clump finding algorithms are ill-suited to our scientific objective. For details on an analysis of cloud properties in the Antennae, see \cite{leroy2016}.

An initial approach was to position a simple Cartesian-like coordinate grid on the moment 0 map and determine the IMR within each pixel. While this certainly met our criteria for an objective method and is thereby less-limited by human bias, it failed on the grounds that it was not adaptive to the variations in CO emission. Algorithmically separating regions of CO emission with differing intensities allows us to limit the effects that result from averaging over regions of differing surface brightness. We decided to adopt a more sophisticated algorithm than a simple Cartesian-like coordinate grid.  Our aim was to define the ``pixels'' more intelligently, i.e. the bounds of the pixels should roughly trace boundaries between differing levels of CO intensity. 

We first tested traditional Voronoi tessellations (VT). This tessellation method is created through a set of generating points (aptly termed ``generators''). The goal of VT is to obtain equal levels of signal-to-noise (S/N, or another measure of emission strength) in a bin. This is done by assigning every pixel, $x_k$, to its nearest generator, $z_i$. In essence, the bins are defined to minimize the quantity $|x_k-z_i|$. By definition, the boundaries between generators are therefore perpendicular bisectors of the connecting segment between the points. Since the boundaries of bins in a VT are straight lines, for complex and diffuse data of $\sim$spherical molecular clouds, the VT does not optimally follow the intensity contours of the CO emission in the data. In previous works utilizing VT, the bin sizes and distribution were made sensitive to density fluctuations by using a bin accretion method. Since we were looking for an algorithm that will follow a curved boundary in detail, we opted to use weighted Voronoi tessellations (WVT).

\subsection{WVT binning algorithm}
\label{sec:wvt}

The WVT binning algorithm was developed for use in sparse X-ray data. It is advantageous over the traditional Voronoi tessellations by introducing an additional parameter, the scalelength $\delta_i$, to be associated with each bin. This scalelength works to stretch or compress the metric inside each bin, which are now defined to minimize the quantity $|x_k-z_i|/\delta_i$. The adaptability of the metric allows for the boundaries of each bin to have curvature that approximately follows a constant gradient of the CO emission intensity. This allows an objective, but adaptive, pixellation of the CO integrated intensity map into bins within which the IMR can be derived.

We made use of the WVT binning algorithm by \citet{DS2006}, which is a generalization of the algorithm developed in \citet{CC2003}. The original algorithm builds the tessellation such that pixels are accreted into bins until a target signal-to-noise (S/N) level is reached. This is appropriate for X-ray data where there exists independent Poisson noise in each pixel. Since the noise in interferometric images is correlated in a more complicated manner, this target parameter is ill-suited for our data. We adapted this algorithm such that pixels are accreted to bins until a limiting luminosity level ($L_{\rm target}$) is reached. Given a conversion prescription from CO radiation luminosity to total mass, by seeking a target total luminosity we are effectively probing a particular gas-mass scale in each tessellation.

Rather than running our adapted WVT binning algorithm over the entire CO moment0-map of the Antennae overlap region, we masked the CO moment0-map to only consider significant (5$\sigma$) star-forming gas emission. The rms of the CO moment0-map was calculated with CASA in isolated patches of the map and was found to be $\sigma \sim 11$\,K\,km/s. We then created a mask of the CO map that only included locations of emission greater than 55\,K\,km/s. The WVT binning algorithm was then run on the masked CO moment0-map for the following limiting luminosity levels:  $3.6\times10^5$\,K\,km/s\,pc$^2$, $7.3\times10^5$\,K\,km/s\,pc$^2$, and $14.5\times10^5$\,K\,km/s\,pc$^2$. The tessellations of differing luminosity levels probe different size and mass scales. Figure \ref{fig:tesselTimeline} shows snapshots of the WVT binning algorithm at work during the 25\%, 50\% and 100\% completion stages.


\subsection{Calculating stellar mass in WVT bins}
\label{sec:wvtstellarmass}

The WVT bin boundaries were applied to the optical image. With the units in $e^{-}$/s, the flux was summed over each WVT bin. Taking the PHOTFLAM keyword from the image header as $7.07236\times 10^{-20}$~ergs/cm$^2$/\AA/$e^{-}$, the $e^{-}$/s per bin were converted to a flux in ergs/cm$^2$/s/\AA~ per bin. This flux was converted to an absolute magnitude assuming an $I$ band flux for Vega of $f_{vega} = 1.134\times10^{-9}$~ergs/cm$^2$/s and a distance to the Antennae of $d=22$~Mpc \citep{schweizer2008}, 
\begin{equation}
M_{I} = -2.5\log\left[\frac{(e^{-}/s)7.07236\times10^{-20}}{f_{vega}}\right]-5\log{d}.
\end{equation}
For each bin that contained at minimum one cluster, the absolute magnitudes could be corrected for extinction using the best estimate of reddening, E(B$-$V), from \cite{whitmore10}. We converted the reddening values to extinctions at $814$\,nm using $A_V = 3.1E(B-V)$ and the relation between $A_{814}$ and $A_V$ determined by \cite{Cardelli1989}:
\begin{equation}
A_{814} \approx 0.6 \, A_V.
\end{equation}

From Figure \ref{fig:mltessel}, it is evident that there exist bins without any identified star clusters, particularly in SGMCs 2 and 3. For those bins, the IMR cannot be calculated, as we lack a suitable method for determining the M/L ratio for the stellar component. For bins with multiple clusters, we follow the procedure discussed in Section \ref{sec:m/l} and assume the M/L ratio of the bin can be approximated by the median of the M/L ratios of the enclosed clusters. 

Using our best estimate for the mass-to-light ratio in a WVT bin (Section \ref{sec:m/l}), the extinction-corrected absolute $I$ band magnitudes were converted to stellar masses in solar units. The stellar mass was combined with the molecular mass in each WVT bin (calculated using the methods of Section \ref{sec:comass}) following Equation \ref{eq:imr} to obtain the IMR in each WVT bin containing at least one stellar cluster.

\subsection{Errors in the IMR}

\begin{figure}
  \centering
  \includegraphics[trim={2.2cm 0cm 0cm 0cm},clip,width=0.47\textwidth]{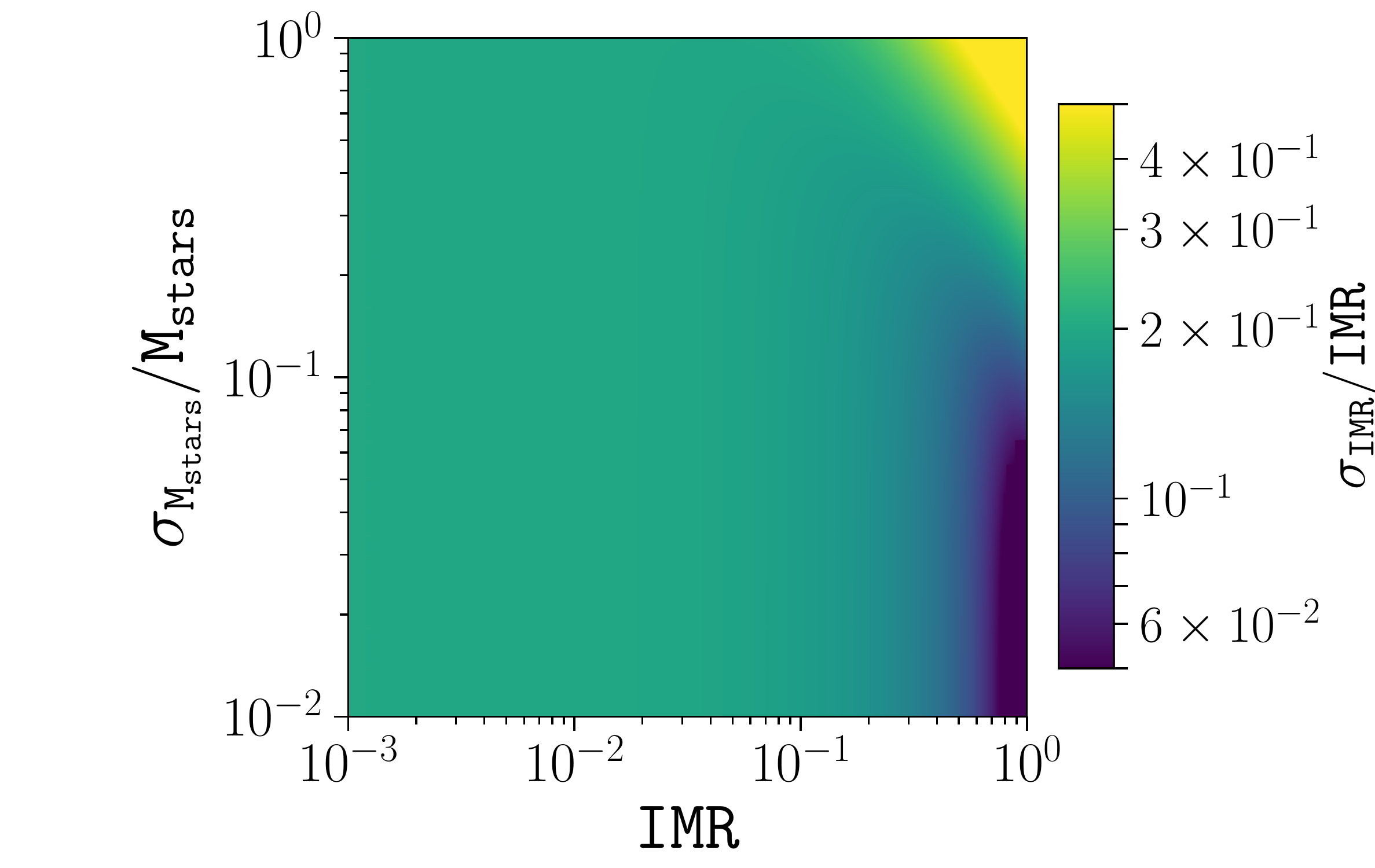}
\caption{The fractional error in the IMR (shown in color-scale) with respect to the IMR and fractional error in the stellar mass component. This assumes an error in the gas mass of 20\%. For much of the parameter space the fractional error in the IMR is dominated by the error in gas mass and hovers around 0.2, which directly reflects the input uncertainty.} \label{fig:IMRerr}
\end{figure}

Complicated and unknown uncertainties in both the stellar and gas mass calculations make estimating formal errors on the resulting IMR nearly impossible. Despite this, it is simple to calculate the dependencies of the error in IMR on several quantities. This allows a better understanding of the areas in parameter space for which the reported values of the IMR$(t)$ can be trusted and to what extent. As an example, we illustrate how the uncertainties depend on the relative errors in the gas mass and stellar mass in Figure~\ref{fig:IMRerr}. Throughout the following calculation, we assume a nominal error in the gas mass of $20\%$, $\sigma_{M_{gas}} = 0.2\,M_{gas}$. Rewriting the IMR from its traditional notation ($M_{stars}/(M_{gas}+M_{stars})$) into a function of the gas-to-stellar mass ratio yields:
\begin{equation}\label{eq:imrerr}
    {\rm IMR} = \left(1+\frac{M_{gas}}{M_{stars}}\right)^{-1}.
\end{equation}
Traditional propagation of error gives the fractional error in the IMR as a function of the IMR itself and the fractional error in stellar mass. Exploring this relationship with varying values of the IMR and the fractional stellar mass error (Figure \ref{fig:IMRerr}) results in fractional errors in the IMR around $\sim20\%$ for much of parameter space. The error in the IMR is minimized as the ratio approaches unity with errors in the stellar mass less than $\sim10\%$.

\section{Results}
\label{sec:results}
\subsection{SGMC Masses and averaged IMRs}
\label{sec:sgmc}

The Antennae galaxies contain molecular clouds of unusually high mass. \cite{wilson2000} derived mass values that are $5-10$ times more massive than the largest molecular complexes of similar size in the grand-spiral galaxy M51 \citep{rand1990}. \cite{wei2012} measured clouds in the overlap region of the Antennae to be larger than any in the Milky Way or nearby disk galaxies \citep{bolatto08,heyer09,solomon87}, causing them to be referred to as Super Giant Molecular Complexes (SGMCs). Within the SGMCs, individual clouds are also grand in scale, \cite{J2015} found a $>5\times10^6\,{\rm M_{\odot}}$ cloud with a radius of $<24\,{\rm pc}$, making it one of the densest clouds known. As some of the most massive molecular complexes, updated molecular mass values for the Antennae SGMCs remain crucial for improved understanding of star-formation in intense environments.

As a consistency check with previous work, we determined the mass of each WVT bin in the three SGMCs following Section \ref{sec:comass} and summed over these values in each of the three SGMCs to derive total gas mass. For the first, second, and third SGMCs, the total gas mass is approximately $7.8\times 10^8$~M$_{\odot}$, $6.3 \times 10^8$~M$_{\odot}$, and $6.6\times 10^8$~M$_{\odot}$. These values are within a factor of $\sim2$ with previous mass estimates of \cite{Wilson2003} for the three SGMCs: $6.3 \times 10^8 {\rm ~M_{\odot}}, 3.9 \times 10^8 {\rm ~M_{\odot}}$, and $9.3 \times 10^8 {~\rm M_{\odot}}$ in SGMCs 1, 2, and 3, respectively. 

The averaged IMRs are 0.2\%, 0.3\%, and 0.3\% for SGMCs 1, 2, and 3, respectively. While these values are quite low, this is reasonable given the enormity of these molecular clouds and the large age range of the stellar clusters enclosed.

\subsection{Star Formation Efficiency from Optical clusters}
\label{sec:sfe_optical}

The instantaneous mass ratio IMR$=\rm M_{stars}/(M_{stars}+M_{gas})$, was calculated for each aperture. The prescription for determining the M/L ratio of a bin, and therefore the corresponding stellar mass, is described in Section \ref{sec:m/l}. The prescription for determining the total gas mass from the CO(3-2) gas emission is described in Section \ref{sec:comass}. These values were then combined to determine the IMR associated with each stellar cluster. We expect that this method will have a bias toward regions with optically visible (un-embedded) clusters that are more efficient at expelling gas. Indeed, \cite{cabrera-ziri15} found that for the three most massive clusters in the overlap region (of ages between 50-200\,Myr), the gas mass is only $<9\%$ of the stellar mass.

\begin{figure}
    \centering
    \includegraphics[width=0.47\textwidth]{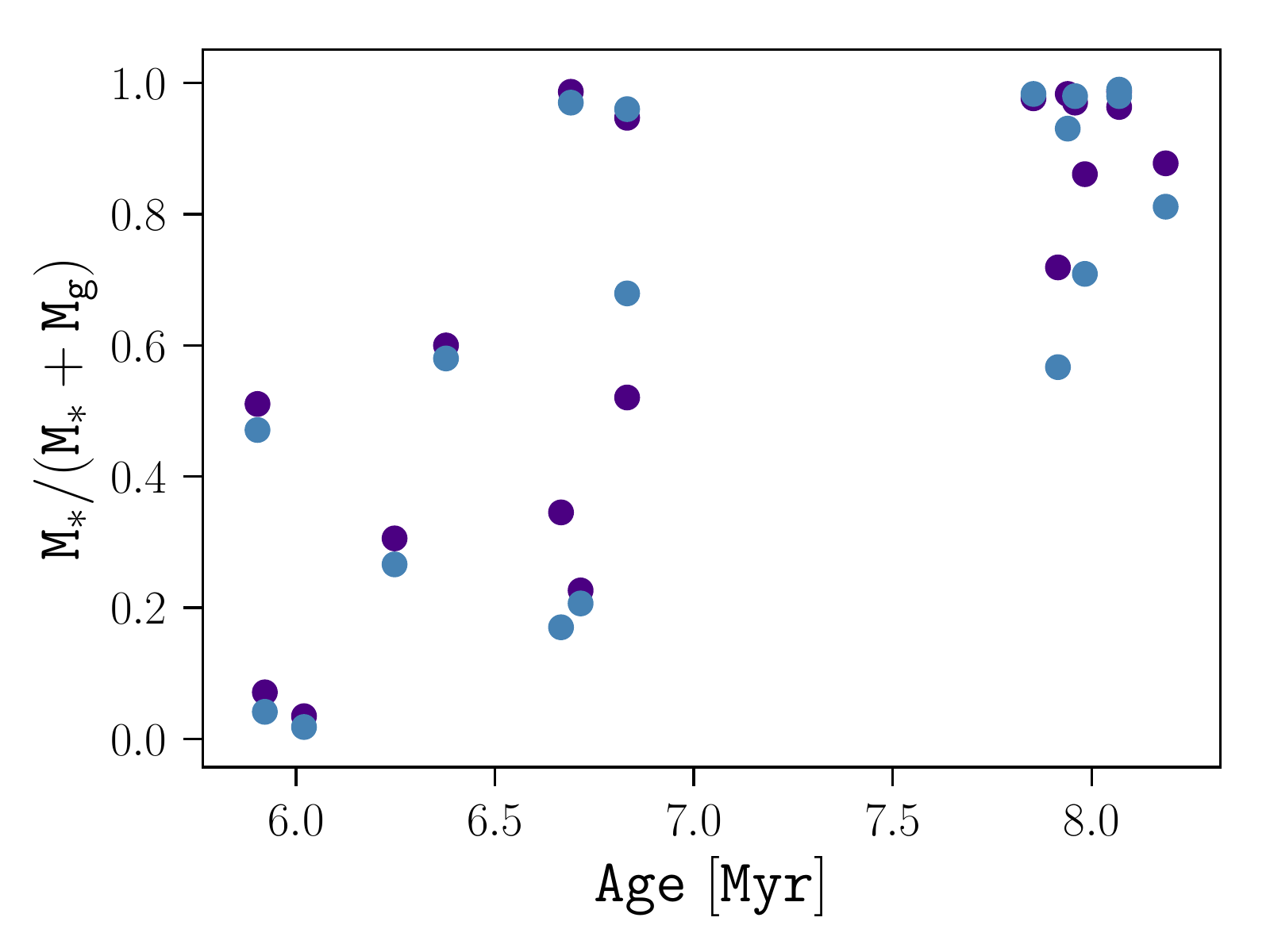}
    \caption{The instantaneous mass ratio (IMR), defined in Section \ref{sec:intro} as a function of time. The IMR was calculated for different aperture sizes around identified optical clusters with the purple and blue points corresponding to an aperture radius of 25 and 50 pc. See Figure \ref{fig:IMRerr} for a description of the errors in the calculated IMR$(t)$ values.}
    \label{fig:sfe_opt}
\end{figure}

The resulting IMR values are presented in Figure~\ref{fig:sfe_opt}. We see a clear trend for the IMR value to increase with cluster age, as we expected, with some regions reaching a value of unity (100\% of the gas has been removed and/or converted to stars) at ages of $<10$~Myr. For ages $\lesssim 2$~Myr (before which we do not expect supernovae and winds from Wolf-Rayet stars to have had a significant effect), the IMR values range from close to zero to $\sim 0.5$. 

A significant caveat to this approach is that for a given optical cluster, we do not know whether the CO emission in the aperture is associated with the cluster, or simply along the same line-of-sight.  Therefore, we expect a fraction of older clusters to {\it appear} to be associated with molecular material, when in fact they are not.  This contamination will result in scatter toward lower IMR values, as discussed in Section~\ref{sec:sfe_co}. The centering of measurements on either the product of star formation (stellar clusters), or the gas produces different results on these small size-scales. \cite{zhang01} found that feedback effects may modify the Schmidt law at scales below 1 kpc. A similar effect of star-formation relations breaking down on small-scales was presented by \cite{schruba10} in M33 and by \cite{kruijssen14} for a sample of regions including the solar neighborhood, the Central Molecular Zone of the Milky Way, a `disk galaxy,' a `dwarf galaxy', and a starbursting `sub-mm galaxy.'

\subsection{Star Formation Efficiency from CO map}
\label{sec:sfe_co}

\begin{figure}
    \centering
    \includegraphics[width=0.47\textwidth]{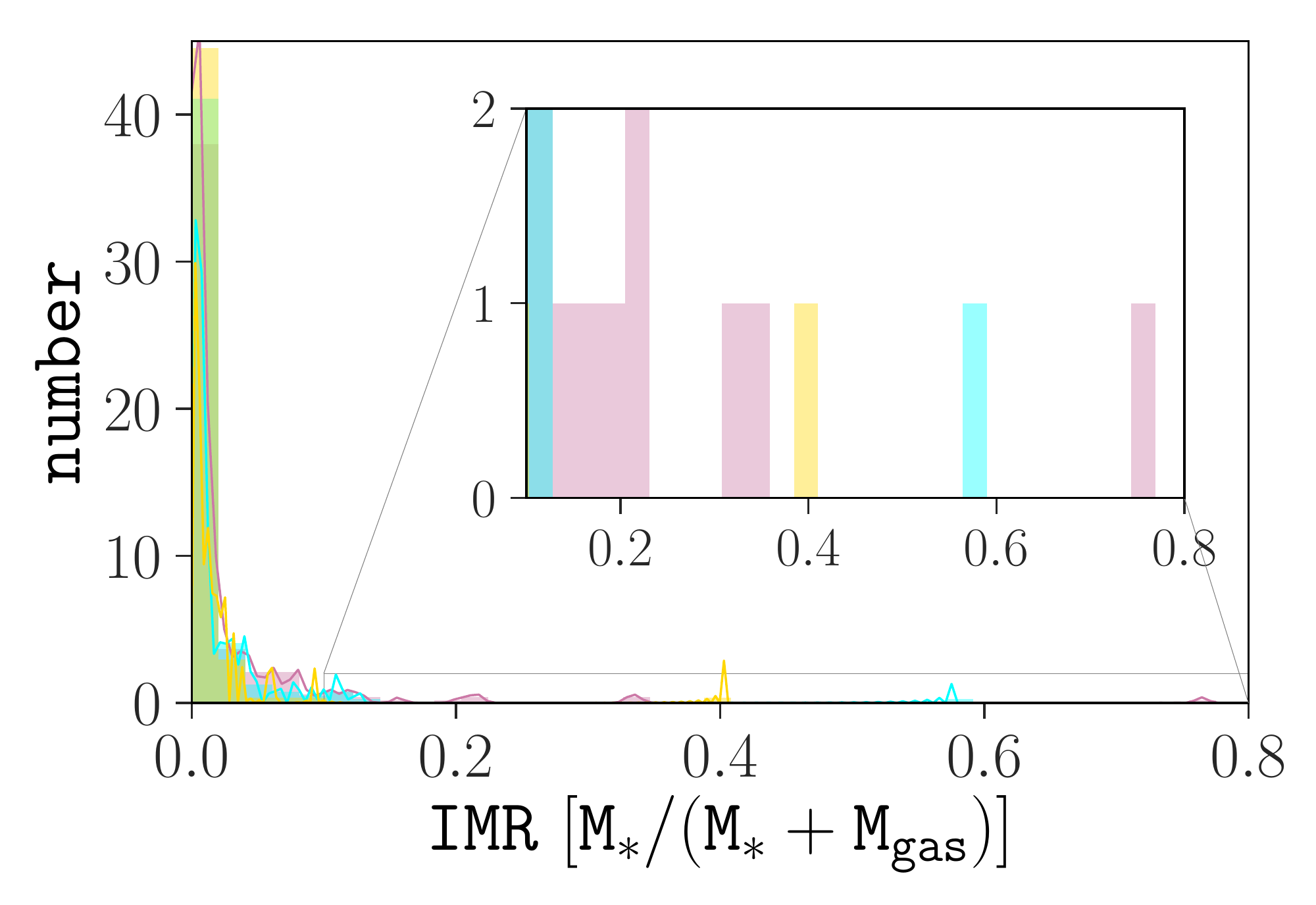}
    \caption{Distribution of the IMR for the various WVT bin sizes: $L_{target}=3.6\times 10^5$ \,K\,km/s\,pc$^2$ (mauve), $L_{target}=7.3\times 10^5$ \,K\,km/s\,pc$^2$  (cyan), and $L_{target}=14.5\times 10^5$ \,K\,km/s\,pc$^2$ (gold).}
    \label{fig:sfe_hist}
\end{figure}

\begin{figure}
\gridline{\fig{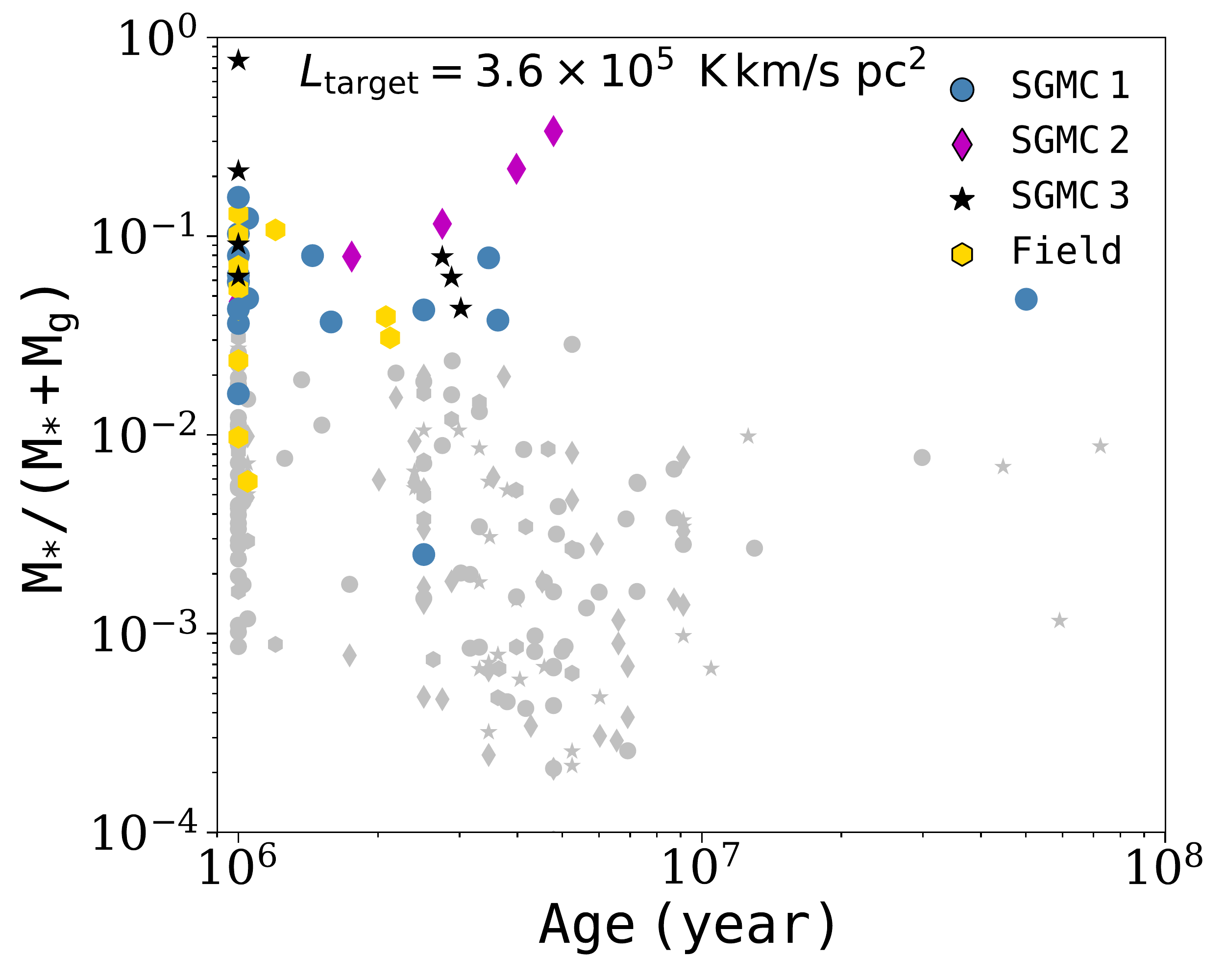}{0.45\textwidth}{}}
\vspace{-10.5mm}
\gridline{\fig{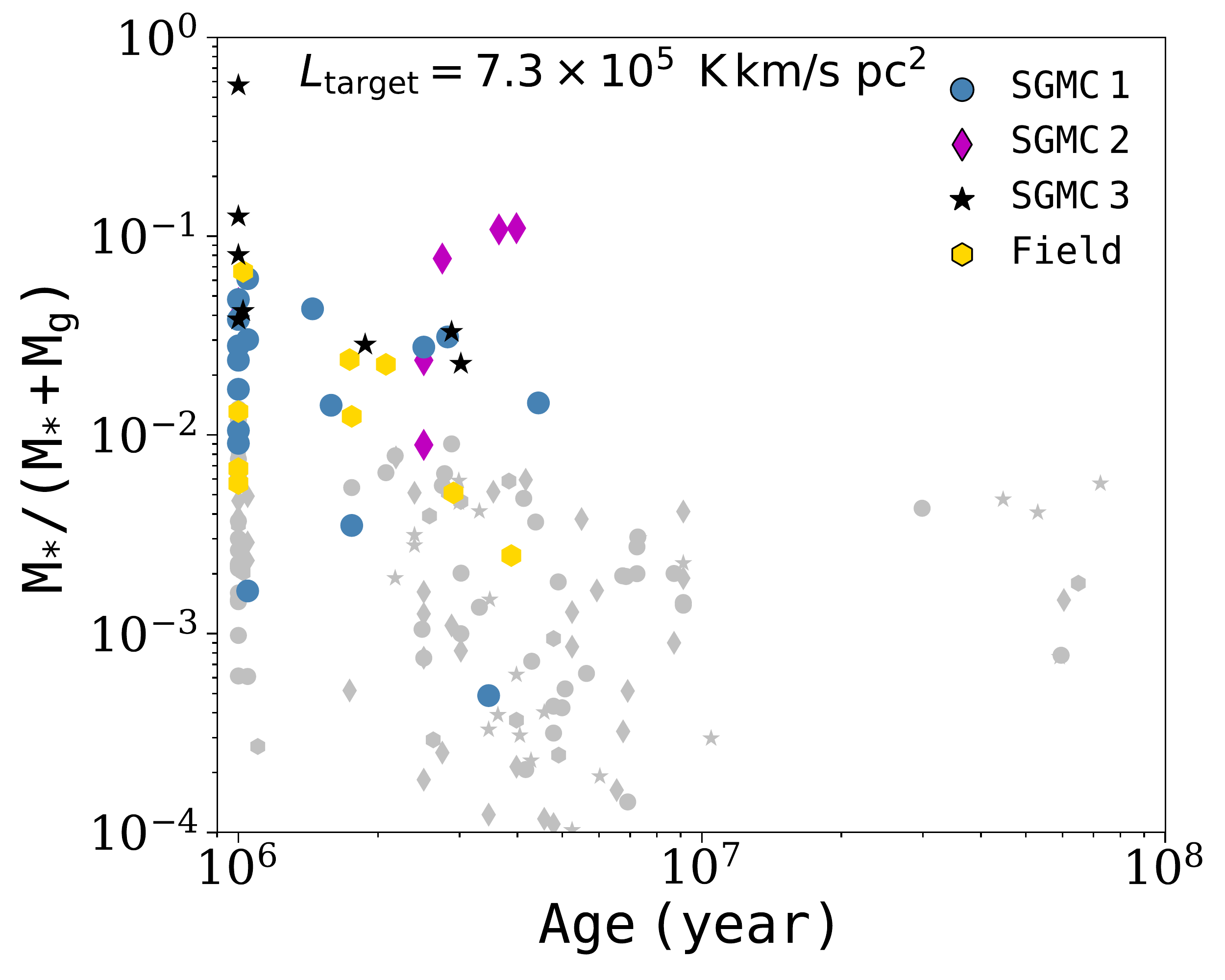}{0.45\textwidth}{}}
\vspace{-10.5mm}
\gridline{\fig{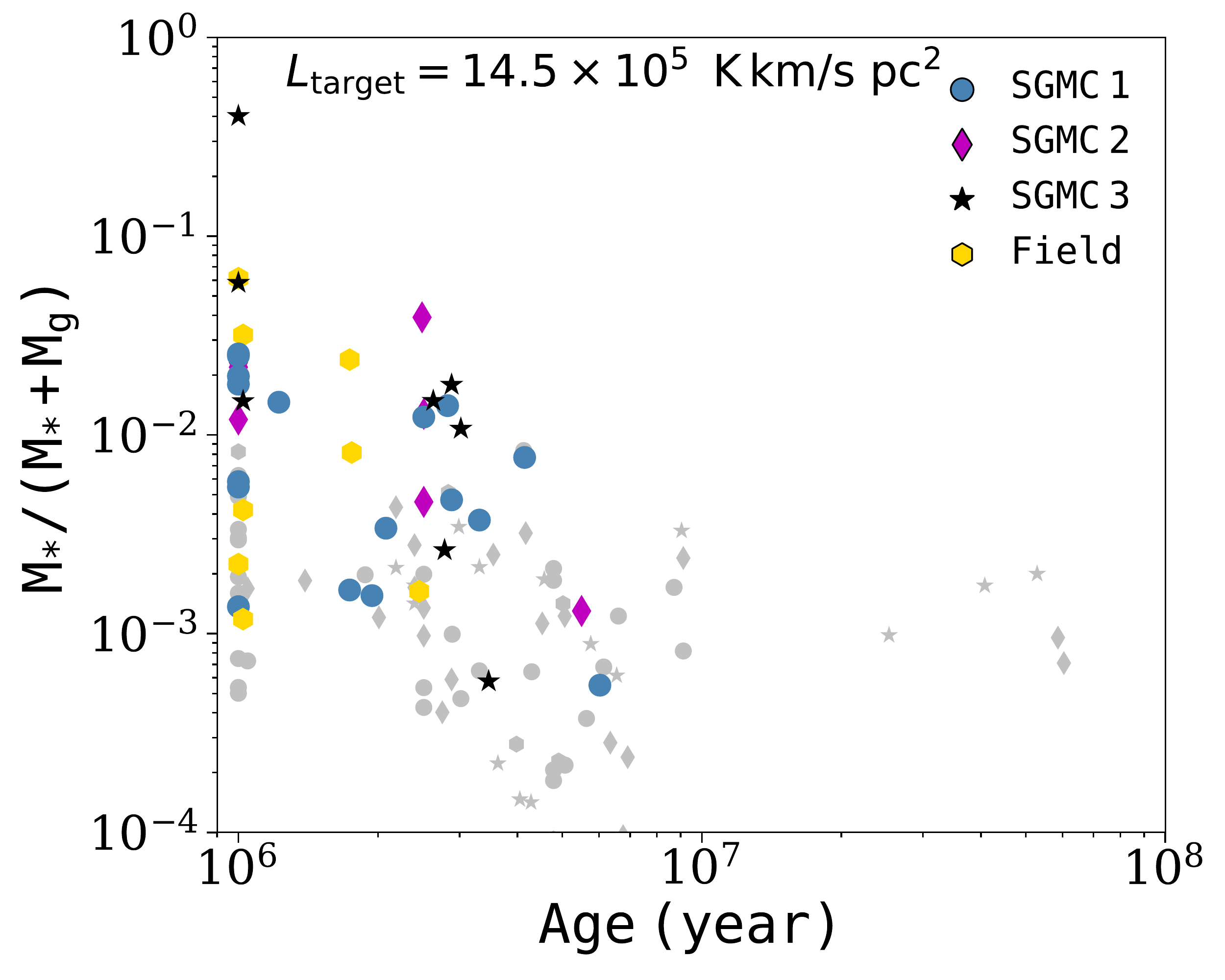}{0.45\textwidth}{}}
\vspace{-6mm}
\caption{The stellar mass to total mass ratio (IMR(t)) is shown as a function of the median cluster age in each WVT bin, with the inclusion criterion for clusters being a concentration index {\it CI} $> 1.52$. This is done for WVT tesselations with target bin CO luminosity of $\sim 3.6\times10^5$\,K\,km/s\,pc$^2$, $\sim 7.3\times10^5$\,K\,km/s\,pc$^2$, and $\sim 14.5\times10^5$\,K\,km/s\,pc$^2$, shown from top to bottom, respectively. Colored points represent WVT bins whose total stellar cluster mass is greater than $10^5\,{\rm M}_{\odot}$, where gray points represent bins with a total stellar cluster mass less than $10^5\,{\rm M}_{\odot}$. See Figure \ref{fig:IMRerr} for a description of the errors in the calculated IMR$(t)$ values.\label{fig:SFEplots}}
\end{figure}

\begin{figure}
\gridline{\fig{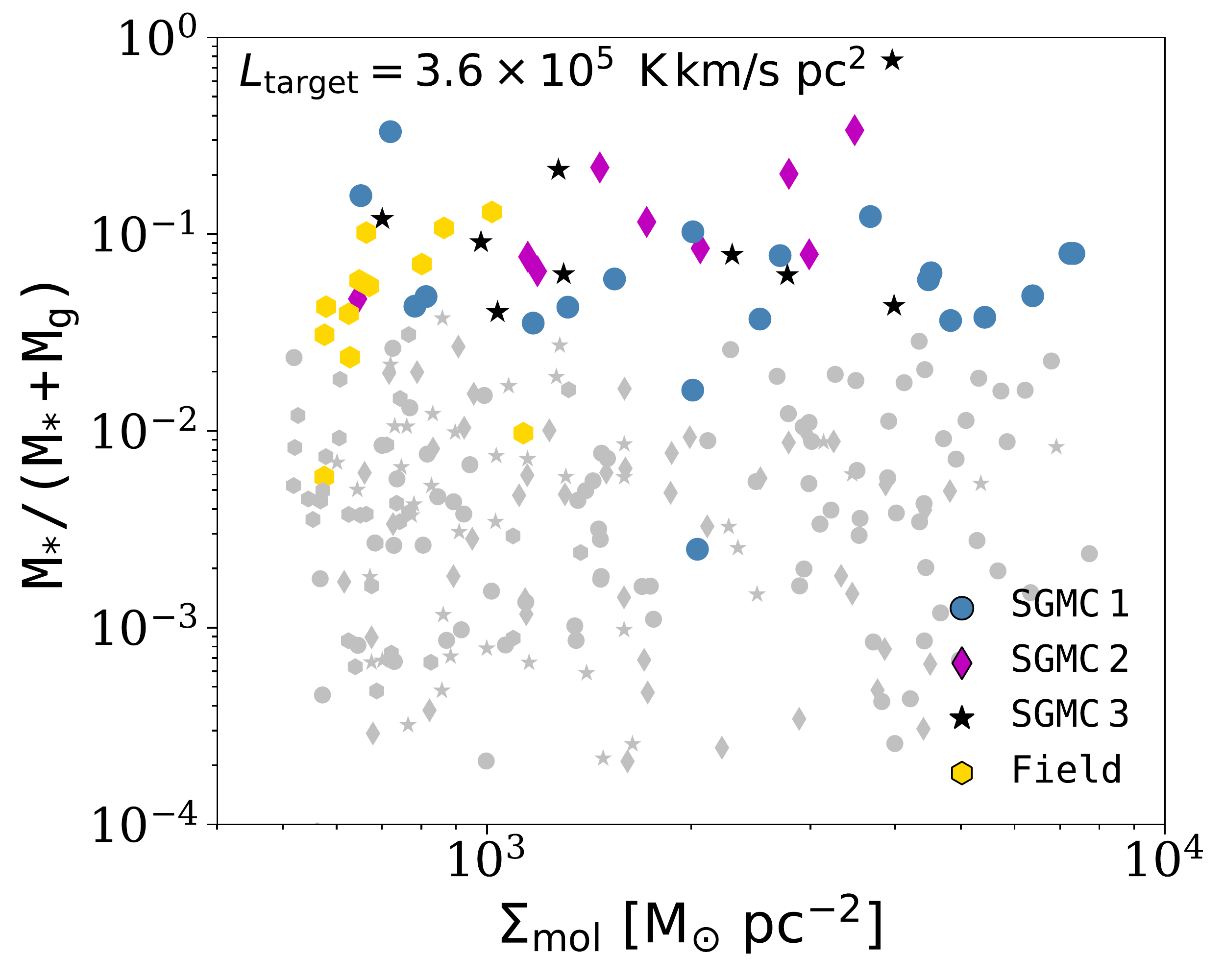}{0.45\textwidth}{}}
\vspace{-10.5mm}
\gridline{\fig{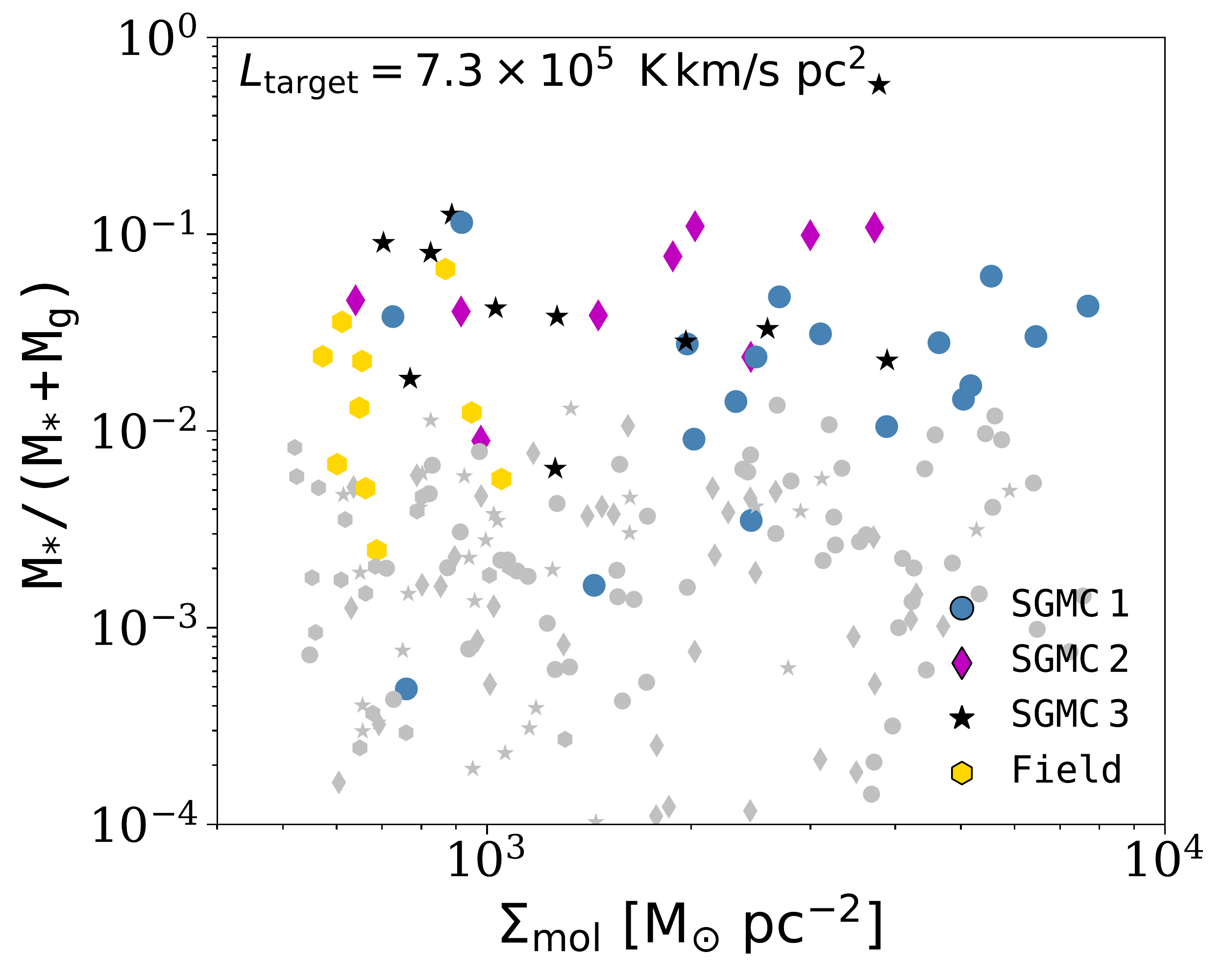}{0.45\textwidth}{}}
\vspace{-10.5mm}
\gridline{\fig{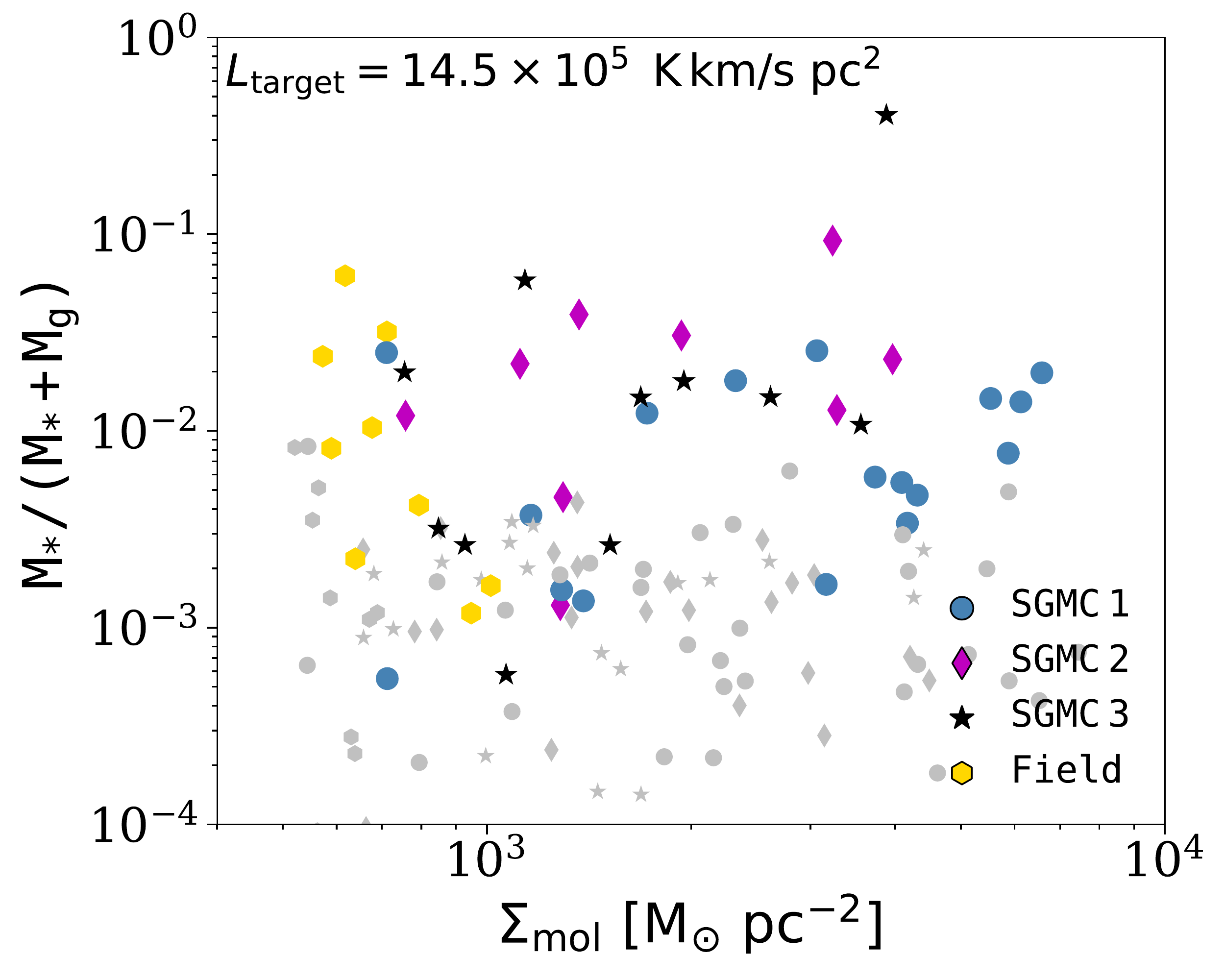}{0.45\textwidth}{}}
\vspace{-6mm}
\caption{The instantaneous stellar mass to total mass ratio (IMR) is shown as a function of the molecular surface density of each WVT bin. This is done for WVT tesselations with target bin CO luminosity of $\sim 3.6\times10^5$\,K\,km/s\,pc$^2$, $\sim 7.3\times10^5$\,K\,km/s\,pc$^2$, and $\sim 14.5\times10^5$\,K\,km/s\,pc$^2$, shown from top to bottom, respectively. Colored points represent WVT bins whose total stellar cluster mass is greater than $10^5\,{\rm M}_{\odot}$, where gray points represent bins with a total stellar cluster mass less than $10^5\,{\rm M}_{\odot}$. See Figure \ref{fig:IMRerr} for a description of the errors in the calculated IMR values.\label{fig:SFE_SBplots}}
\end{figure}

The instantaneous mass ratio (IMR) was also calculated for all WVT bins that had an associated stellar cluster (see Section \ref{sec:co}). This procedure was completed for WVTs resulting from target luminosities of $L_{target}$ = $3.6\times10^5$\,K\,km/s\,pc$^2$, $7.3\times10^5$\,K\,kms\,pc$^2$, and $14.5\times10^5$\,K\,km/s\,pc$^2$. We find very few WVT bins with IMRs greater than $\sim 10\%$, with the mean of the distribution being $2.4\,\%$ (see Figure \ref{fig:sfe_hist}). The cluster with consistently the highest IMR (shown as a black star in Figures \ref{fig:SFEplots}, \ref{fig:SFE_SBplots}, \ref{fig:SFE_AIplots}) is identified as WS80 and has been studied extensively in the literature \citep[e.g.][]{whitmore02}. As the size of the WVT bin increases, the fractional increase of stellar light due to both the diffuse stellar component and compact clusters within the WVT bin increases more slowly than the increase in the diffuse CO(3-2) emission. Correspondingly, as the size of the WVT bins ($L_{target}$) decreases, the IMR values systematically increase. We find that determining the IMR by centering measurements on molecular gas does not follow the predicted trend as seen in Section \ref{sec:sfe_optical}.

A necessary condition for determining the IMR in a WVT bin is the presence of at least one identified stellar cluster in the projected area of the bin. The age dating of these clusters, done by \cite{whitmore10}, enables us to compare the relationship of the IMR with time, see Figure \ref{fig:SFEplots}. The majority of WVT bins contain only one identified cluster, but in the case where there are multiple, the median age of the enclosed clusters is taken. Our original hypothesis was that as a cluster ages, more of its gas is converted into stars, destroyed, or expelled, and the IMR would approach 100\%. We do not observe this trend in the overlap region of the Antennae galaxies for regions selected based on their CO(3-2) emission. Rather, we observe systematically decreasing IMRs for clusters of increasing age. This result is interesting, but may be better explained by an observational bias rather than an astrophysical phenomenon (this bias is explored further in Section \ref{sec:bias}). 
In particular, because the bins are determined based on their CO(3-2) flux (for which a 5$\sigma$ detection is required), we expect that this method will be systematically biased toward regions with significant CO emission. By requiring the presence of significant CO emission in the analyzed WVT bins, we bias ourselves exclusively towards clusters with a significant mass of gas. We repeated all procedures using a 2$\sigma$ and 3$\sigma$ threshold for the CO emission to determine whether the trend of decreasing IMR with age would change, but no significant change was observed. Given that the clusters are also identified in {\it projection}, we also expect there to be contamination from clusters simply along the line-of-sight, and not physically associated with the molecular gas, which would systematically manifest itself as older clusters appearing to be gas-rich.  These combined effects may explain the systematically low IMR values for relatively older clusters in the overlap region of the Antennae galaxies. 

\subsection{Comparison of methods for an arbitrary cluster}
\label{sec:cluser_comp}

\begin{figure}
  \centering
  \includegraphics[trim={0.8cm 2.2cm 0cm 0cm},clip, width=0.47\textwidth]{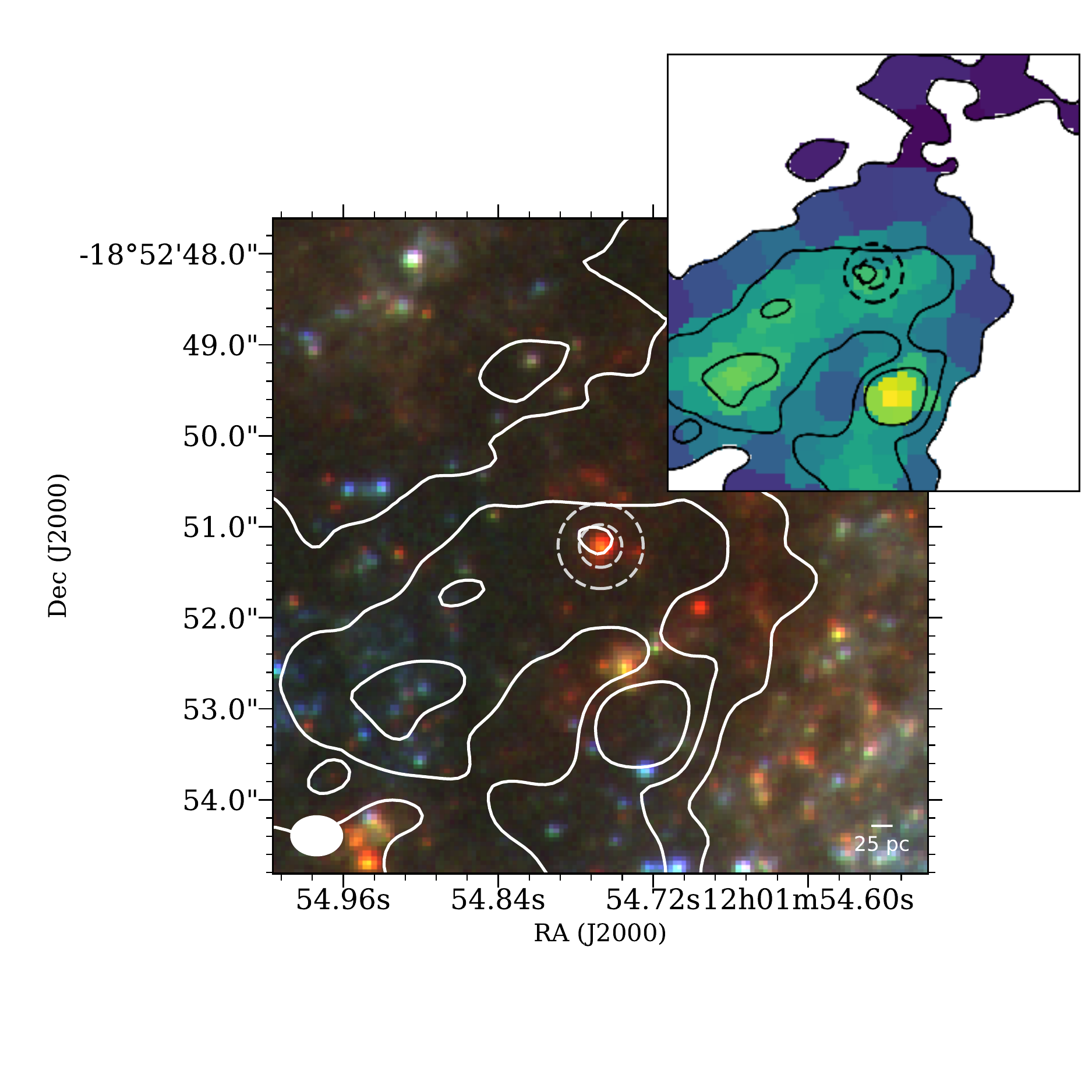}
\caption{An optical image of SGMC 2 centered on the 19th most massive cluster identified in \cite{whitmore10}. White contours represent CO(3-2) emission with levels [55,220,440]\,K\,km/s. Dashed gray lines represent 25\,pc and 50\,pc circular apertures used to determine the IMR based on locations of stellar clusters. The beam of the ALMA Cycle 0 observations is shown in the bottom left.  The inset on the upper right shows the tessellation with a target luminosity $L_{\rm target}$ of $7.3\times10^5$\,K\,km\,s$^{-1}$ of the same region with identical contours and apertures overlaid.} \label{fig:clusterZoom}
\end{figure}

As a direct comparison of the two methods, we take an arbitrary cluster that is both one of the ``50 Most Massive'' clusters (included in the analysis of Section \ref{sec:stellar}) and surrounded by a 5$\sigma$ detection of CO(3-2) emission (included in the analysis of Section \ref{sec:co}). The cluster chosen is the 19th most massive cluster in the entire Antennae system (Figure \ref{fig:clusterZoom}), with an ID of 18848 in \cite{whitmore10} and nominal values of the stellar mass and age estimate of $1.2\times10^6\,{\rm M_{\odot}}$ and 4.8\,Myr, respectively.

Centering the measurement on the stellar cluster yields an IMR of 23\% and 21\% for the 25\,pc and 50\,pc apertures, respectively. In comparison, the WVT bins containing this cluster have IMRs of 34\%, 11\%, and 10\% for increasing $L_{target}$, respectively. In order to compare these values, we consider the physical area of the circular apertures and WVT bins. The 25\,pc and 50\,pc radii apertures have areas 1960\,pc$^2$ and 7850\,pc$^2$, respectively, while the WVT bins have areas of 874\,pc$^2$, 2040\,pc$^2$, and 4440\,pc$^2$, for increasing $L_{target}$. The smallest WVT bin has an IMR greater than the apertures. This is reasonable due to its small size, which has a higher fraction of cluster light to diffuse stellar light and minimizes the mass from the gas reservoir that encompasses the cluster. The larger IMR ratios for the apertures centered on the stellar clusters are likely due to the differing focuses of the two methods. While the circular apertures are centered on the cluster the WVT bins follow the gas emission, which is not symmetrically distributed and leads to larger gas masses and lower IMRs.

\subsection{Dependence of IMR on surface brightness and extinction}
\label{sec:IMRvsOther}


The WVT binning algorithm allows us to investigate the dependence of the IMR on various physical parameters. For example, we can explore whether the efficiency of star formation depends on the surface density of the CO gas in molecular clouds. The ratio of the flux in a given WVT bin and the area in units of pc$^2$ probes the CO gas density. Since our WVT bins were defined algorithmically from the data itself, we are able to objectively measure the dependence of the IMR as a function of molecular gas surface density. When considering WVT bins that include optical clusters, the IMR does not seem to correlate with the CO surface brightness (Figure \ref{fig:SFE_SBplots}), which agrees with results of \cite{leroy17} showing that the small-scale surface density of clouds has a relatively weak effect on star formation. This is rather unexpected, and may be truly physical or it may indicate that the surface density is not a good probe of the volumetric density of CO gas and we are therefore unable to establish a relationship between IMR and CO gas density. Even when considering WVT bins with only the more massive optical clusters of ${\rm M}>10^5$~M$_\odot$, the relationship between IMR and surface brightness remains largely consistent with a random distribution. (Figure \ref{fig:SFE_SBplots}).

\begin{figure}
\gridline{\fig{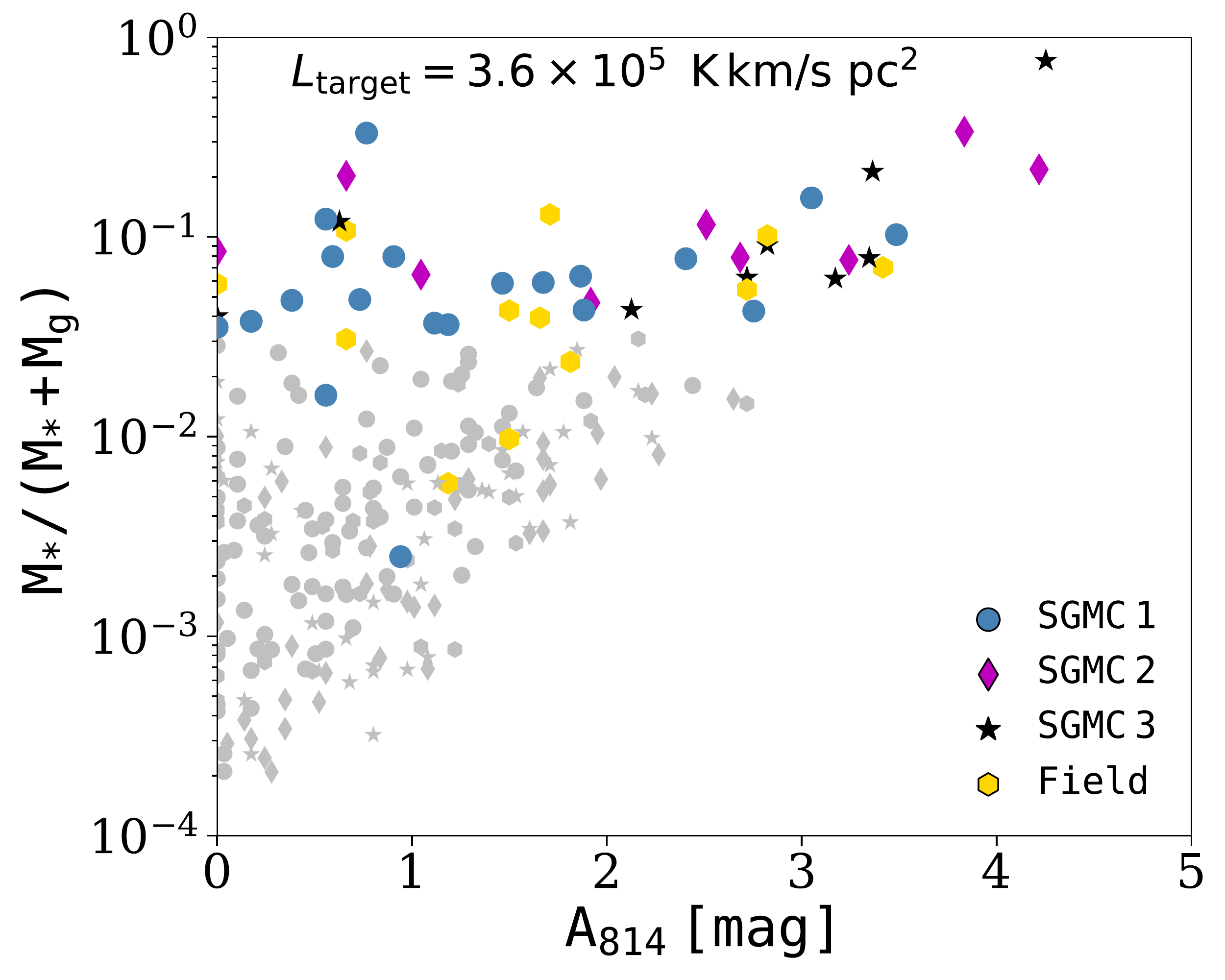}{0.45\textwidth}{}}
\vspace{-10.5mm}
\gridline{\fig{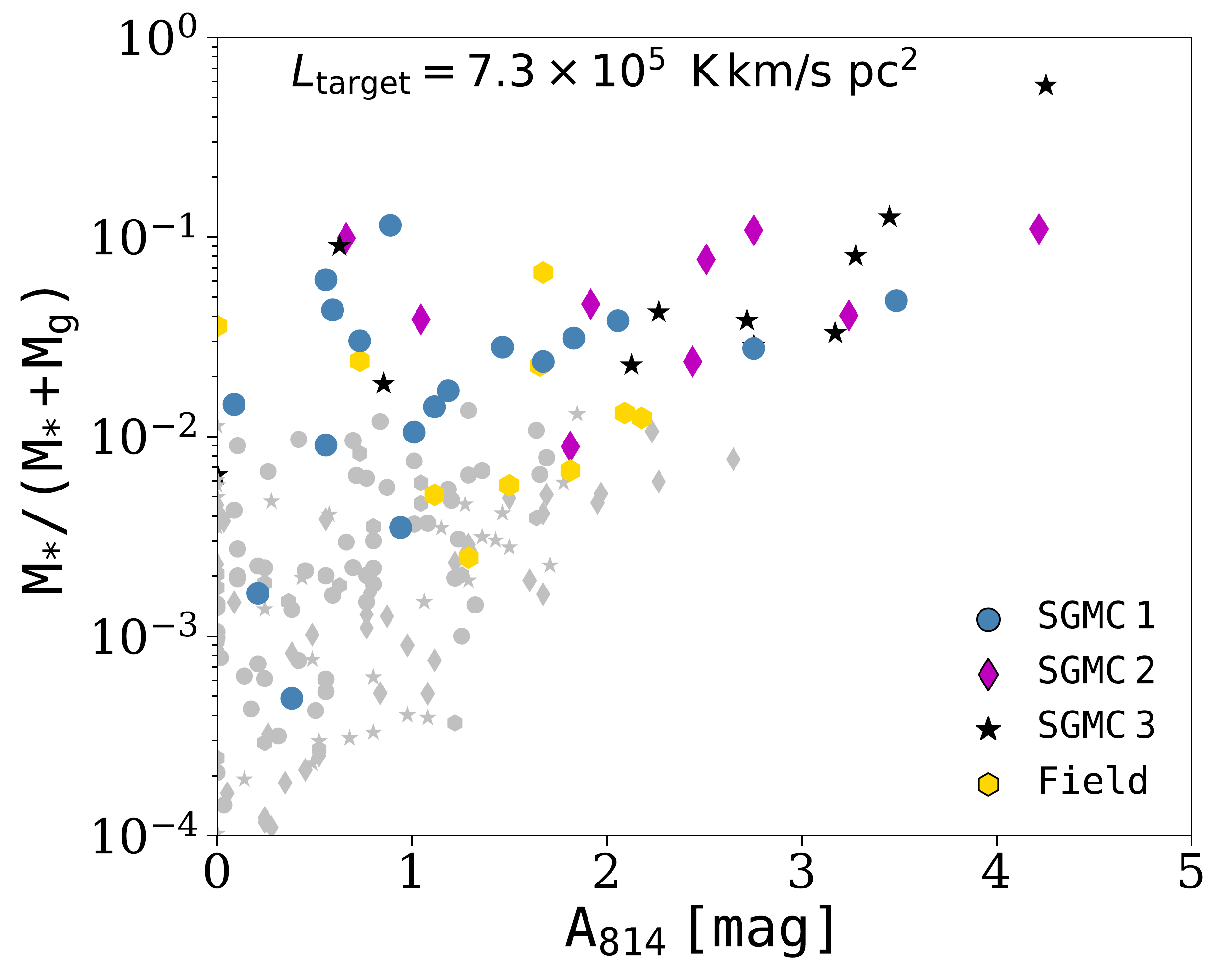}{0.45\textwidth}{}}
\vspace{-10.5mm}
\gridline{\fig{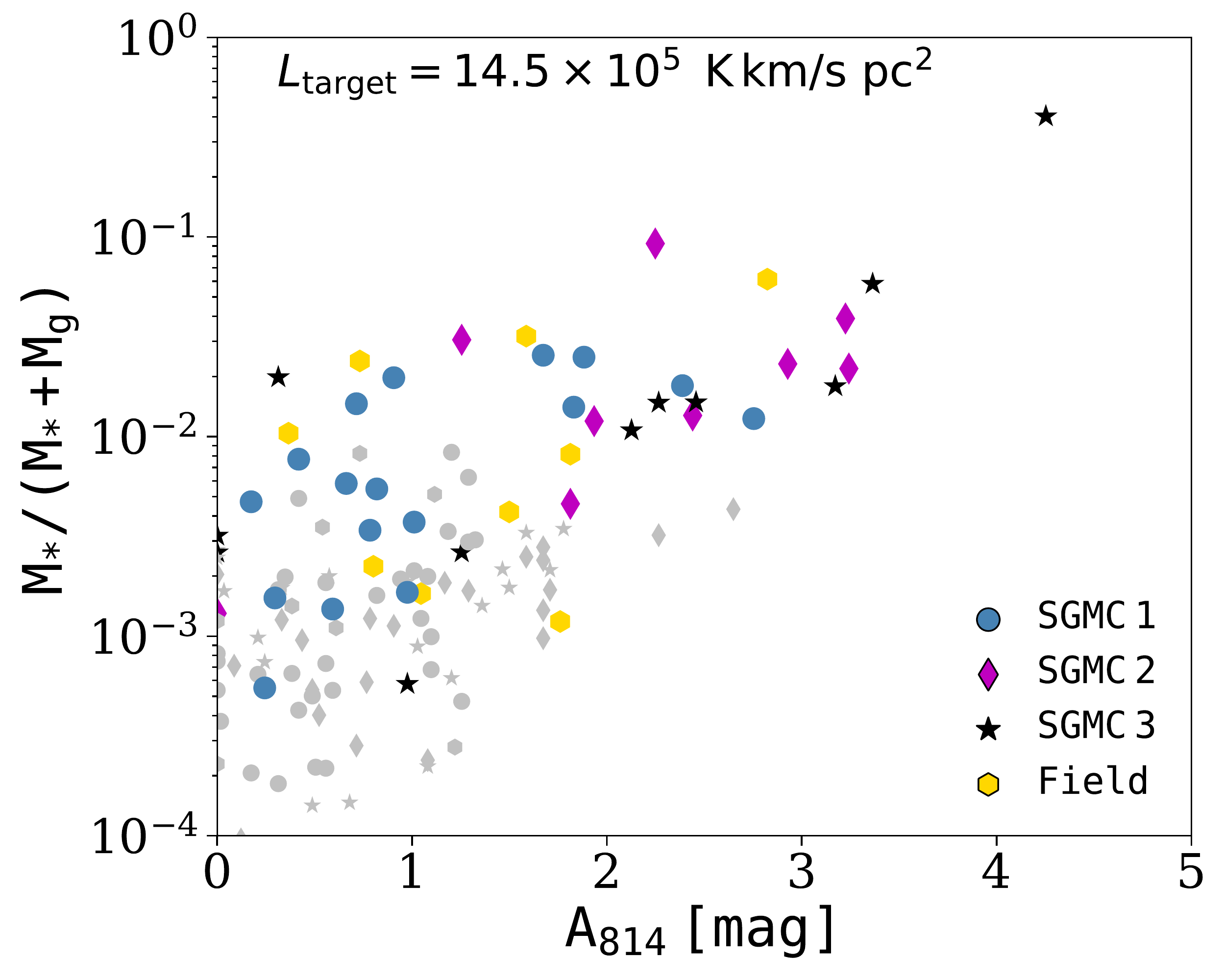}{0.45\textwidth}{}}
\vspace{-6mm}
\caption{The instantaneous stellar mass to total mass ratio (IMR) is shown as a function of the median extinction at 814\,nm for all clusters in each WVT bin. This is done for WVT tesselations with target bin CO luminosity of $\sim 3.6\times10^5$\,K\,km/s\,pc$^2$, $\sim 7.3\times10^5$\,K\,km/s\,pc$^2$, and $\sim 14.5\times10^5$\,K\,km/s\,pc$^2$, shown from top to bottom, respectively. Colored points represent WVT bins whose total stellar cluster mass is greater than $10^5\,{\rm M}_{\odot}$, where gray points represent bins with a total stellar cluster mass less than $10^5\,{\rm M}_{\odot}$. See Figure \ref{fig:IMRerr} for a description of the errors in the calculated IMR values.\label{fig:SFE_AIplots}}
\end{figure}

\begin{figure*}
\gridline{\fig{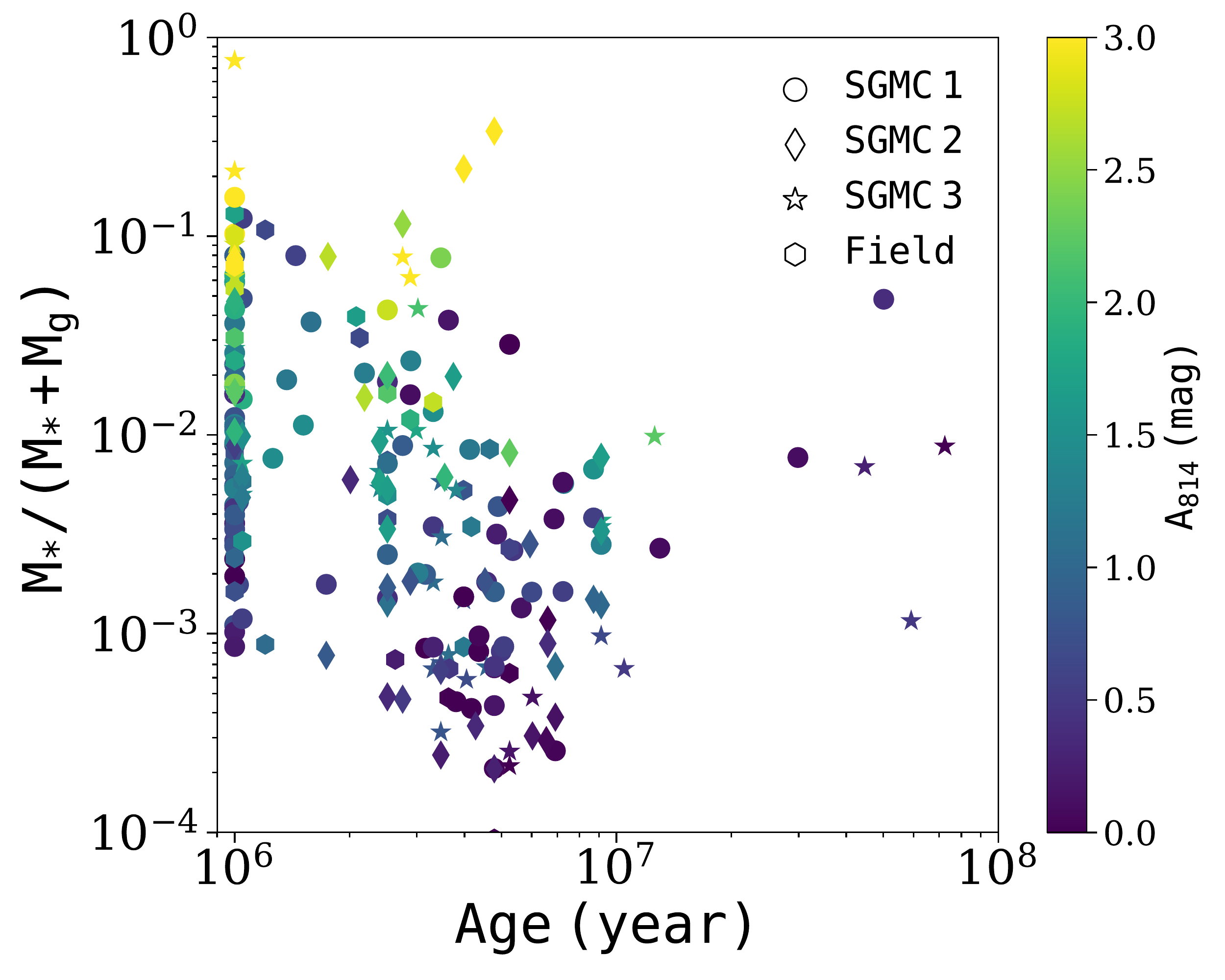}{0.47\textwidth}{}
\fig{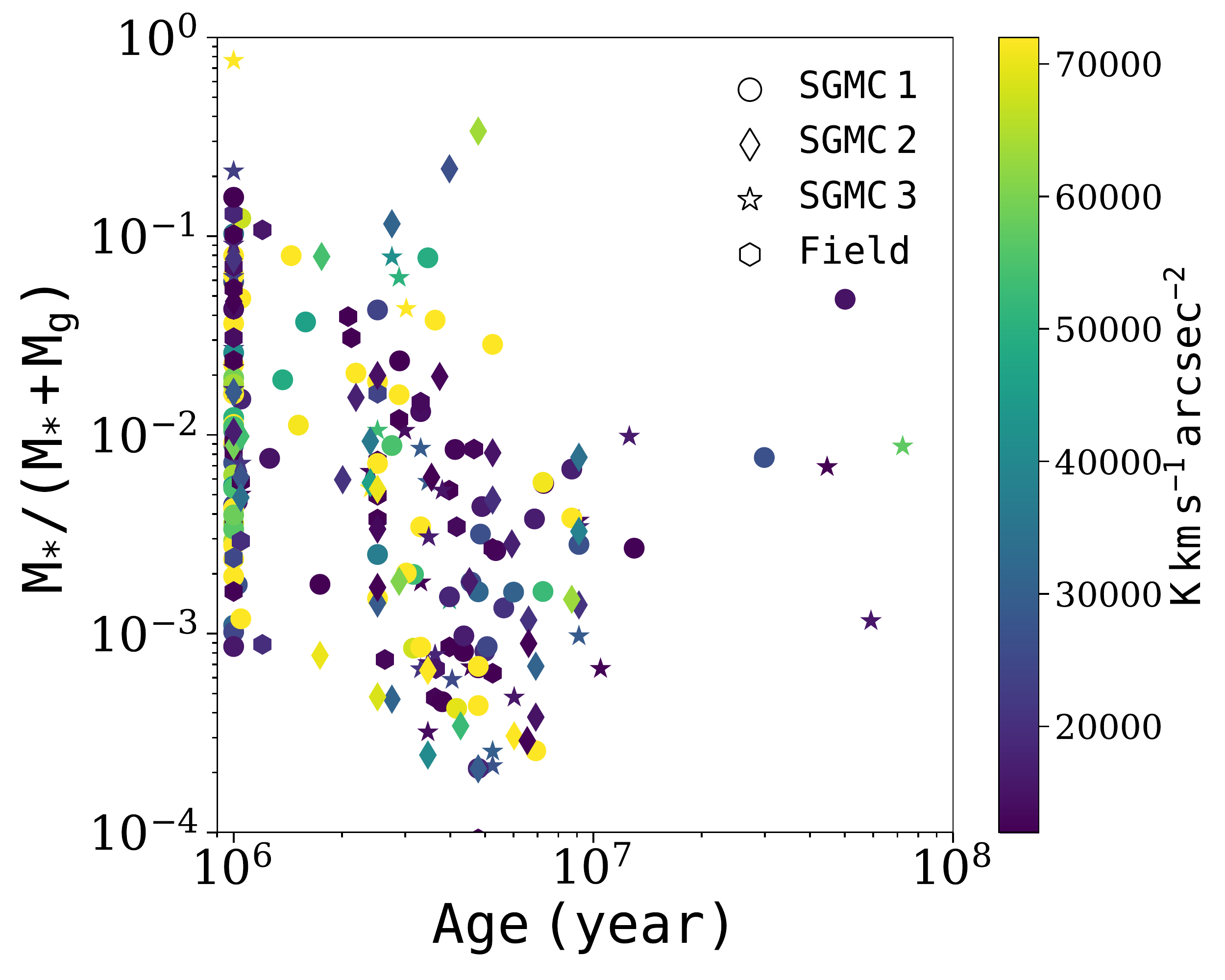}{0.47\textwidth}{}}
\vspace{-4.5mm}
\gridline{\fig{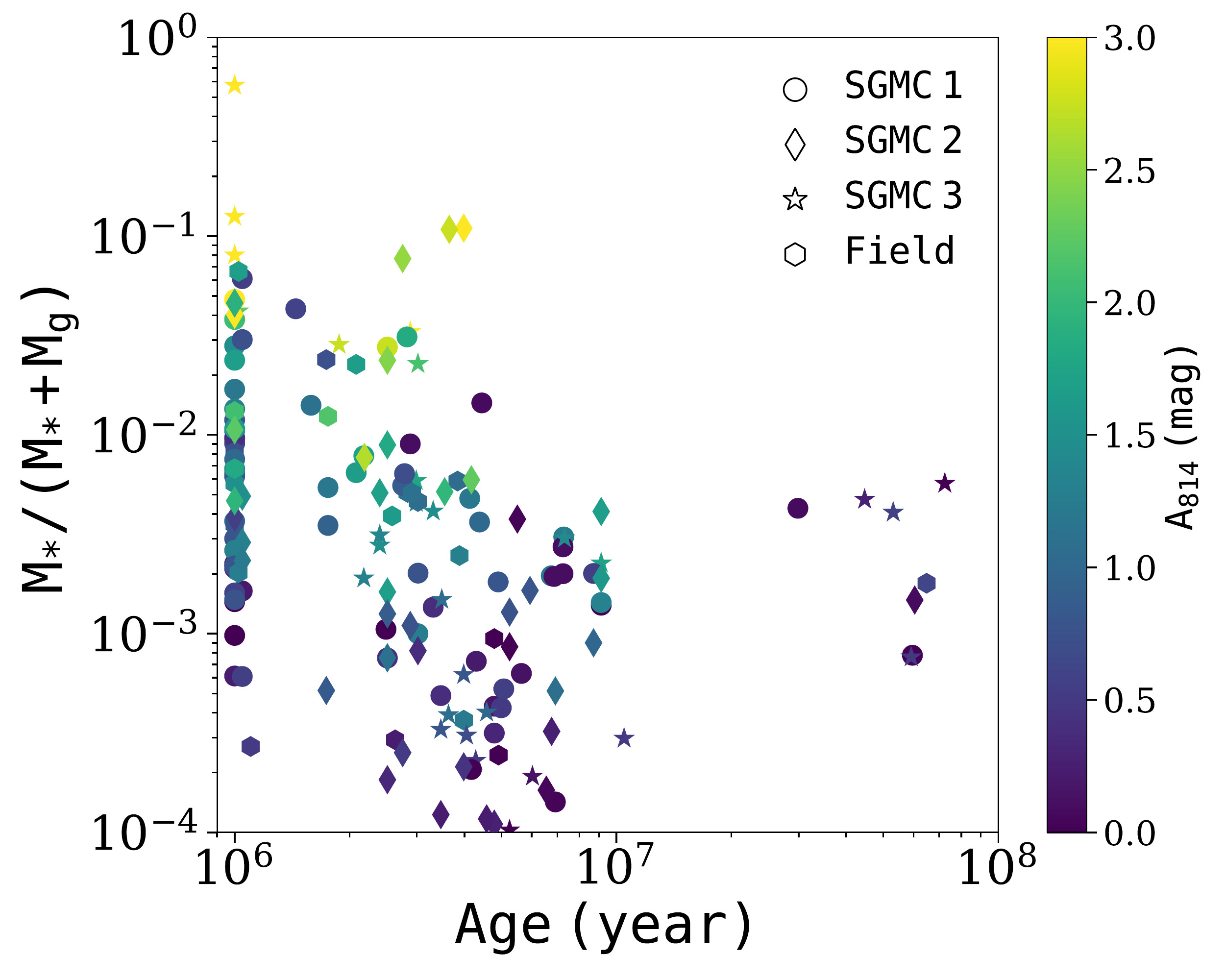}{0.47\textwidth}{}
\fig{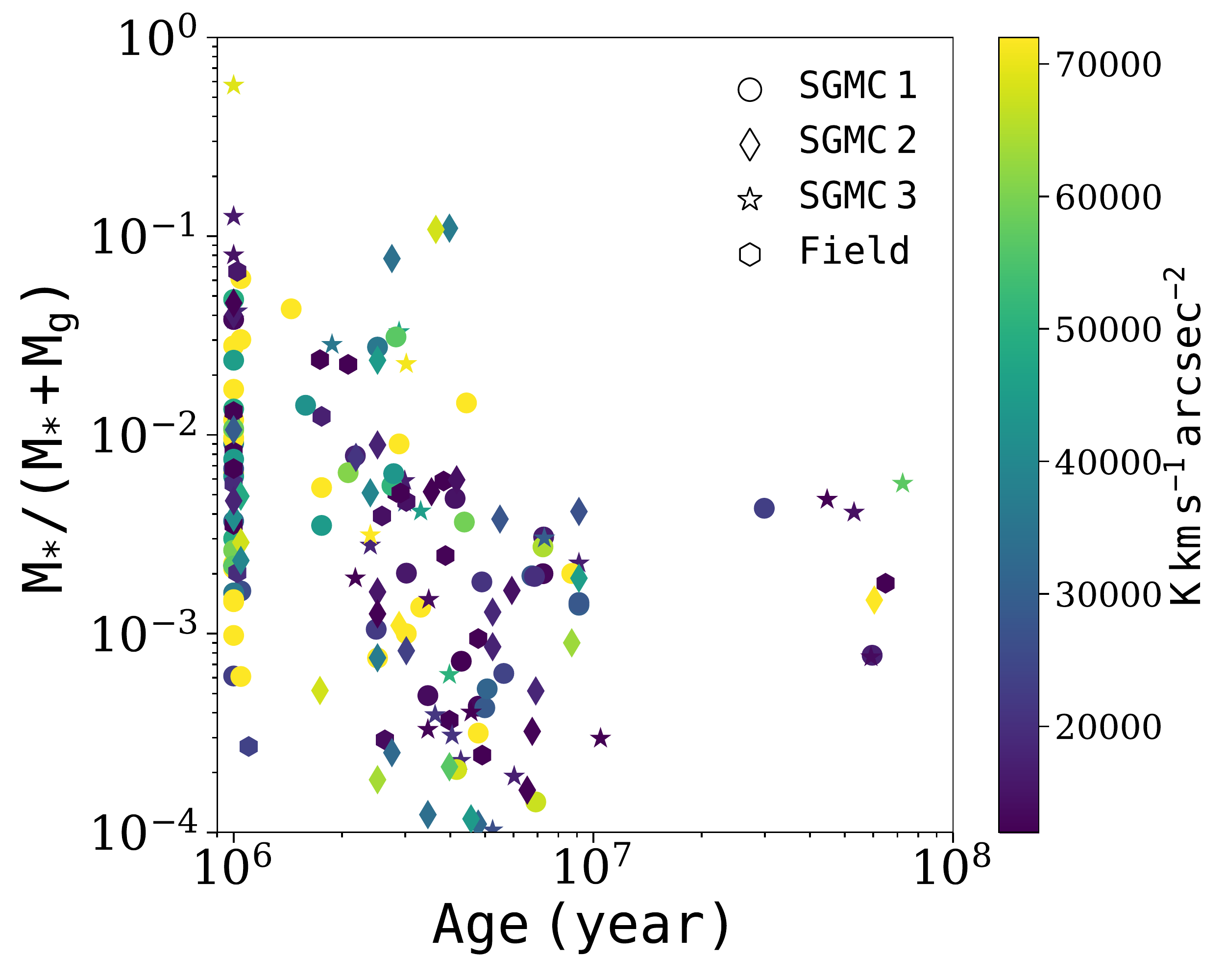}{0.47\textwidth}{}}
\vspace{-5mm}
\gridline{\fig{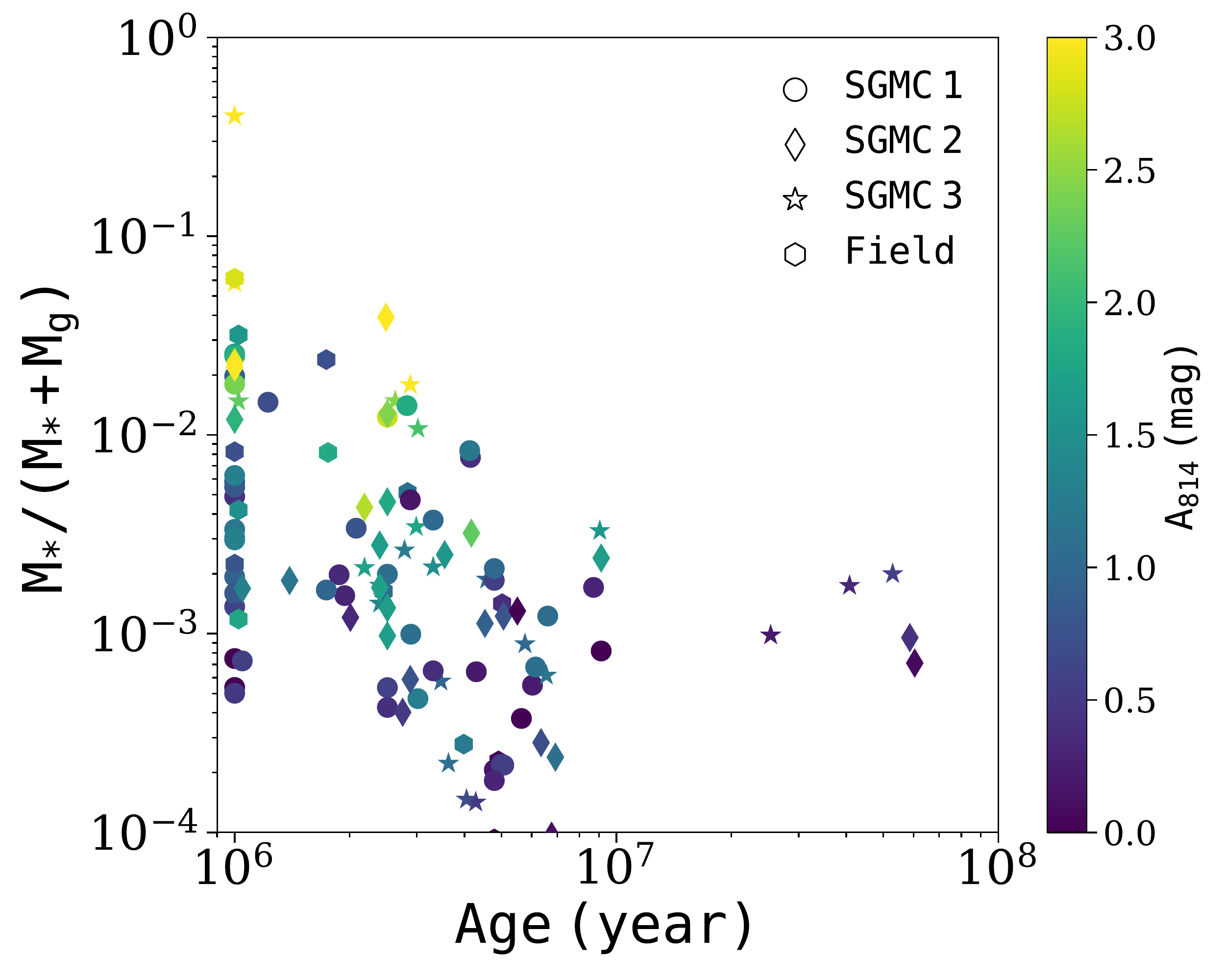}{0.47\textwidth}{}\fig{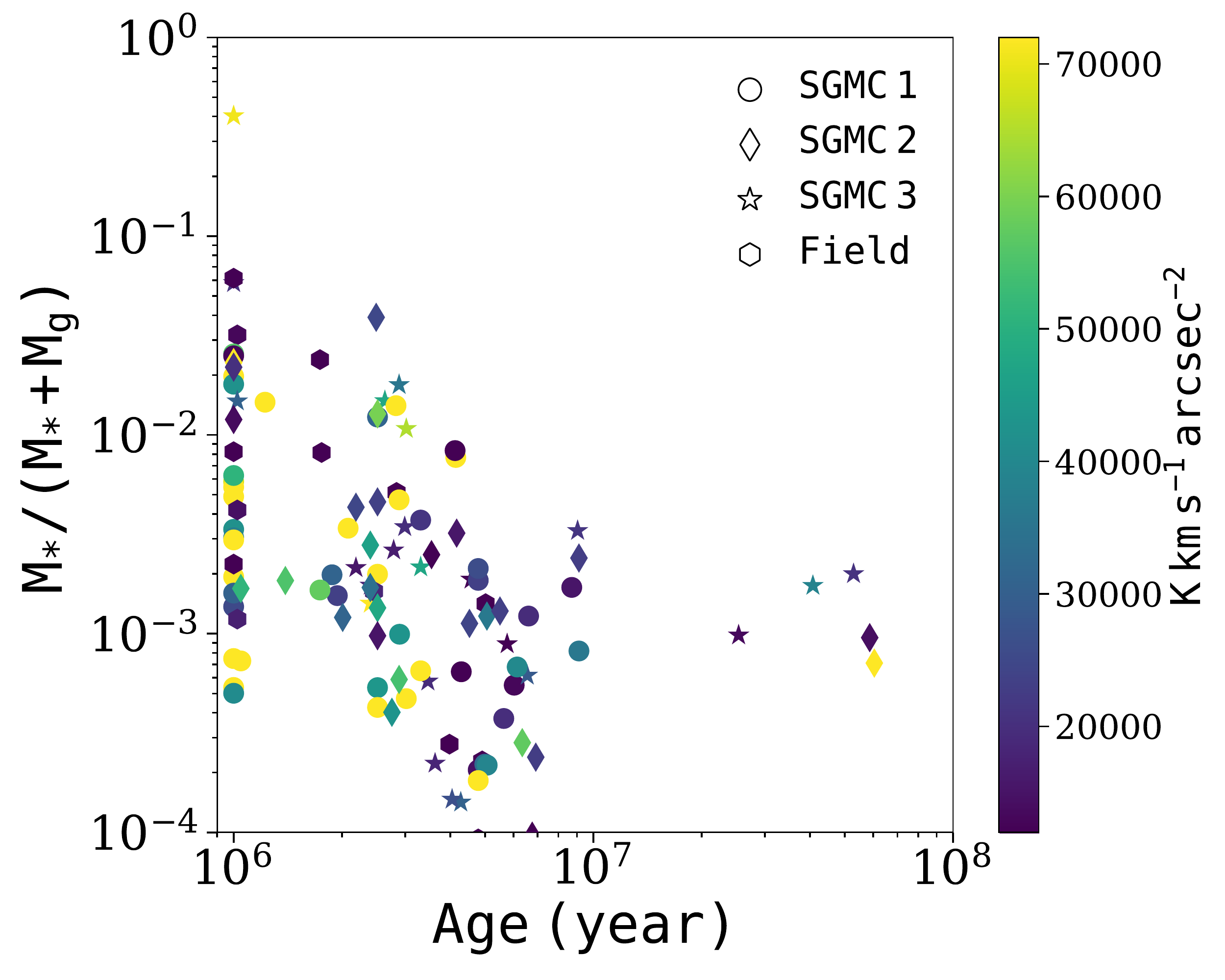}{0.47\textwidth}{}}
\vspace{-3mm}
\caption{The stellar mass to total mass ratio (IMR) is shown as a function of cluster age, with the inclusion criterion for clusters being {\it C} $> 1.52$, for a WVT bin luminosity of $3.6 \times 10^5$\,K\,km/s\,pc$^2$, $7.3 \times 10^5$\,K\,km/s\,pc$^2$, and $14.5 \times 10^5$\,K\,km/s\,pc$^2$ from top to bottom, respectively. Left: The colors of individual points correspond to the median extinction, $A_I$, for the clusters in the bins taken from \cite{whitmore10}. Right: The colors of individual points correspond to a measure of the surface brightness of each bin. See Figure \ref{fig:IMRerr} for a description of the errors in the calculated IMR$(t)$ values. \label{fig:SFE_AISBplots}}
\end{figure*}

In addition to the relationship between the IMR and CO surface brightness, we can also investigate the dependency of IMR and the $814$\,nm $A_{814}$ extinction in each of the WVT bins. Higher extinction values may indicate that the cluster is more deeply embedded in the molecular cloud from which its stars are forming. Due to the increase in the amount of available dense gas in this environment, we expect higher IMR for clusters with higher values of extinction. Since the WVT bins were determined independent of extinction, they provide an objective means for investigating whether more deeply embedded clusters indeed correspond to higher IMRs. We see a clear trend across all SGMCs of increasing IMR with extinction, confirming that more deeply embedded clusters are capable of higher rates of SFE (Figure \ref{fig:SFE_AIplots}). A comparison of the IMR with respect to both age and extinction can be seen in Figure \ref{fig:SFE_AISBplots}, along with a comparison of the IMR to both age and surface brightness.

\section{Discussion}
\label{sec:discussion}
\subsection{Biases of the two methods}
\label{sec:bias}

Each of the two primary methods we used to estimate the IMR(t) for regions in the Antennae is subject to limitations.  With the WVT method, there is power in mathematically defining regions whose areas adjust based on the amount of CO emission at a given point.  However, it constrains our analysis to regions where there is significant CO emission, which inherently biases our results toward regions rich in molecular gas, but which may not have associated star formation or the star formation episode may be in its early stages. For these young clusters, it is likely that we are missing optical light and thereby underestimating the stellar mass associated with the WVT bin. This method also selects and highlights slightly older clusters that are either inefficiently producing stars or feedback from massive stars has failed to clear out much of the initial gas from the molecular cloud. These scenarios are expected to be rare and cannot explain the presence of multiple clusters with ages $\gtrsim5$\,Myr with very low IMRs. The more plausible explanation is that these regions are ``contaminated'' by clusters along a WVT bin's line-of-sight and which are not physically associated with the observed molecular gas. This limitation of the algorithm biases our results towards lower values of the IMR, particularly for older clusters.

Because of these biases towards detecting the population of older clusters that {\it appear} to remain in a dense molecular cloud, the results of the evolution of IMR with age by centering our measurement on molecular gas contradict our initial hypothesis. Specifically, we observe a decreasing trend in the relationship between these two quantities.

On the other hand, the trend is reversed for the analysis done by focusing on circular apertures around luminous and massive clusters. This result follows our prediction: the ratio of stellar to total mass increases as stars form and gas is expelled through stellar feedback. Although this result supports our original hypothesis, this method is also subject to biases. Building a sample of clusters by requiring their luminosity and mass to be high will preferentially select a sample of clusters that have been successful at clearing much of their gas reservoir. \cite{whitmore10} found that $\sim15\%$ of clusters in the entire Antennae system were completely dust-obscured. This ratio may be higher in the overlap region due to the larger gas reservoir. While there are still some of the optically brightest clusters visible within the SGMCs, the majority of this sample resides outside of these giant gas clouds. There are very few older clusters that are surrounded by a significant amount of gas, and it is more likely that older clusters with low IMR are a result of chance superposition or incorrect age estimates. This sways the stellar to total mass ratios of this sample to approach unity as age increases.

\subsection{The Observed Instantaneous Mass Ratios and Cluster Survival}
\label{sec:survival}

The ability of star clusters to remain bound and evolve into globular clusters is strongly tied to their SFE. However, different theoretical efforts have not converged on consistent values, but instead have predicted a large range of SFEs necessary for clusters to remain bound. For example, the simplest case without considering environmental effects (e.g. turbulence and non-instantaneous timescales \cite{hills80} requires a SFE of at least 90\%, while \cite{baumgardt07} and \cite{Pelupessy2012} report values as low as $5-10\%$ provided the gas is removed over long timescales. With only $\sim 8$ clusters presented here with IMR $> 10\%$, our findings suggest that SFEs of several tens of percent (but much less than 90\%) are not especially common. Taken at face value, the apparently low SFE with which clusters in the Antennae are forming could contribute significantly their destruction.

Our initial hypothesis that the IMR(t) value should exhibit a trend in which IMR values increase with age and approach unity appears to be approximately borne out (as shown in  Fig.~\ref{fig:sfe_opt}), which is based on localized emission in the vicinity of optical SSCs.  However, this hypothesis is generally inconsistent with the results from the second method based on the intensity of the CO emission (as shown in Fig.~\ref{fig:SFEplots}). These apparently contradictory results can be explained by both line-of-sight contamination and the Antennae system not being a steadily evolving post-starburst, but rather having a significant CO reservoir that is actively forming star clusters.  In a system crowded with both newly-formed and mature star clusters, any arbitrary and sufficiently large aperture is likely to contain a mixed population, which makes it challenging to disentangle the gas associated with any particular cluster of a given age.

For these reasons, we assume that the IMR is better determined through the method that localizes emission associated with optical clusters. If we assume that the IMR roughly equivalent to the SFE for clusters with sufficiently young ages (e.g. IMR$(t\leq10^{6.5})$), we can assess the fraction of clusters that could remain bound.
Based on the analysis discussed in Section~\ref{sec:sfe_optical}, only 2/5 of the youngest massive clusters ($t \leq 10^{6.5}$ years, mass $\gtrsim 7\times 10^5$~M$_\odot$) optically visible clusters considered here appear to have instantaneous mass ratios $\gtrsim 0.5$.  If SFEs as low as $\sim 30$\% could still produce a bound cluster, the relative fraction of these massive clusters that could survive increases to 3/5. Although we recognize small sample statistics may be at play, this fraction does not appear to increase until the SFE threshold is lowered to 5\%, which may only result in bound clusters under very specific physical scenarios (see Section~\ref{sec:1.1}). However, we note that this method inherently assumes that the gas within the 25~pc or 50~pc radius around each cluster is associated with that cluster.  For the youngest clusters, the IMR values for the 25~pc and 50~pc radii are very similar, suggesting that the gas is highly concentrated on the clusters.  However, if the gas associated with the clusters has been over-estimated, the corresponding IMR value will be higher -- suggesting a higher fraction of potentially bound clusters.

If instead we look to the IMR$(t)$ values determined based on the CO emission in Section~\ref{sec:sfe_co}, only a single region appears to have IMR(t$< 10^{6.5}$) $\gtrsim 0.3$, suggesting that very few of the regions at large in this system will produce bound clusters. Thus, on the surface, these results suggest that only a few of the massive clusters formed recently in the Antennae have the chance to evolve into globular clusters, especially considering the number of other physical processes that will affect their survival after the first few million years \citep[e.g.][]{gnedin97, hollyhead15}. This is consistent with other studies \citep[e.g.][]{whitmore04, fall05, whitmore07} that find that only a few percent of young star clusters survive to become old globular clusters.

\subsection{Clearing Timescales for the CO Gas}
\label{sec:timescales}

The timescales on which molecular material is destroyed or expelled from a massive star cluster is important for determining whether these massive clusters could potentially undergo a ``second generation'' of star formation, as has become an outstanding issue in globular cluster evolution in recent years.  Observations of globular clusters have unambiguously shown evidence for multiple chemical populations, however the origin of these populations remains unclear \citep[e.g. Bastian et al. in prep, ][]{cabrera-ziri15}.  

From Section~\ref{sec:sfe_optical}, we can estimate the timescales on which the CO gas is destroyed or expelled from the cluster environments.  Figure~\ref{fig:sfe_opt} shows a clear trend indicating that the relative fraction of CO associated with clusters decreases with age, as expected.  By ages of $\sim 10^{6.7}$~years, some of the clusters appear to have already cleared the molecular material from within radii of 50\,pc (where \cite{whitmore14} found that the lifetime of GMCs in the region is $\approx10^7$ years).  By ages of $\sim 10^{7.5}$~years, this is true for the majority of clusters. This agrees well with the clearing timescales of 2\,Myr and 7\,Myr for gas that is associated and unassociated with the stellar cluster, respectively (Grasha et al., in prep). Moreover, we note that for clusters older than $\sim 10^{6.7}$~years, the IMR values derived from the 25~pc and 50~pc radii are more discrepant (see Fig.~\ref{fig:sfe_opt}), potentially indicating that gas is no longer as centrally concentrated on the clusters by this time.
We hypothesize that the clusters with ages $> 10^{7.5}$~years that appear to have molecular material still associated with them may well simply be contaminated by CO emission along the line-of-sight.

\subsection{The Need for Thermal Radio Observations}

Determination of the IMR through centering the measurement on gas or clusters are both subject to the additional issue that we are likely missing stellar light from the youngest clusters, which may not be visible at optical wavelengths. This additional effect is more strongly felt by the WVT binning method as the resulting bins are systematically more enshrouded in obscuring material than the clusters assembled by having a high luminosity and/or mass. In both methods (but particularly for the one based on CO emission) we are very likely underestimating the stellar mass of the youngest clusters (and older enshrouded ones). This effect extends to the IMR, which is also likely to be slightly underestimated at the youngest ages. Properly accounting for stellar mass of young clusters could be done with high spatial resolution thermal radio observations as an extinction-free tracer of recent star formation. Although \cite{neff00} present VLA images of the Antennae at $6$ and $4$\,cm of resolutions $\sim1''$ and $2\farcs6$, respectively, these observations likely suffer from too much non-thermal synchrotron contamination.



\section{Conclusions}
\label{sec:conclusions}
This work sought to derive the SFE of clusters in the Antennae galaxies, the nearest major-merger to the Milky Way. Age information for the $>10,000$ identified clusters in the overlap region \citep{whitmore10} allowed further investigation into the relationship between the instantaneous mass ratio (IMR) as a function of time, where IMR$(0)$ = SFE. 

We followed two approaches in deriving the IMR$(t)$. The first (Section \ref{sec:sfe_optical}) analyzed the enclosed gas and stellar mass within circular apertures of 25 and 50\,pc. This resulted in a monotonically increasing trend of IMR with age, as it is biased towards clusters capable of ejecting its molecular gas as internal feedback mechanisms turn on.

The second approach (Section \ref{sec:sfe_co}) employed an algorithm to intelligently pixelate the moment-0 CO map such that each bin contains approximately the same flux, defined by the parameter $L_{target}$. This method resulted in a monotonically decreasing trend of IMR with age, as it is biased towards clusters whose feedback was incapable of expelling its molecular gas. The combination of the two methods yields an honest picture of the efficiency of star formation in the Antennae galaxies, from which we conclude:

\begin{itemize}

\item Only a handful of the bins and/or clusters have IMR$(0)$ = SFE $>$ 0.2. On the surface, this implies that the fraction of star clusters that remain bound is small. However, the optical identification of clusters and determination of their masses will be systematically biased toward low apparent stellar masses due to dust, and we are likely underestimating the IMRs. 

\item{Some clusters appear to have removed virtually all of the molecular material at ages as young 10$^{6.7}$ years.  The majority of clusters appear to have removed their molecular material by ages 10$^{7.5}$ years.}

\item There is no observed dependence of IMR$(t)$ on surface density, but a strong dependence on extinction. The dependence of the IMR$(t)$ on extinction can be explained if extinction is acting as a proxy for the evolutionary state of clusters.
The lack of dependence on the surface density of the molecular clouds indicates that the surface density is a poor tracer of the IMR and suggests the value of the IMR is almost independent of the surface density. 


\end{itemize}

\acknowledgments
We thank the referee for many useful comments that improved the paper. This material is based upon work supported by the National Science Foundation Graduate Research Fellowship under Grant No. DDGE-1315231. This research is supported by NSF grant 1413231 (PI: K. Johnson). This paper makes use of the following ALMA data: ADS/JAO.ALMA\#2011.0.00876.S ALMA is a partnership of ESO (representing its member states), NSF (USA) and NINS (Japan), together with NRC (Canada), NSC and ASIAA (Taiwan), and KASI (Republic of Korea), in cooperation with the Republic of Chile. The Joint ALMA Observatory is operated by ESO, AUI/NRAO, and NAOJ. The National Radio Astronomy Observatory is a facility of the National Science Foundation operated under cooperative agreement by Associated Universities, Inc.

\vspace{5mm}

\bibliographystyle{apj}
\bibliography{antennaerefs}




\end{document}